\documentclass[aps,pre,showpacs,twocolumn]{revtex4}

\usepackage[T1]{fontenc}
\usepackage{amssymb}
\usepackage{amsbsy}
\usepackage{amsmath}
\usepackage{graphicx}
\usepackage{hyperref} 

\begin{document}

%\graphicspath{{./figures/}}

% AUTHORS' MACROS
\def\kin{k_{\rm in}} 
\def\beq{\begin{equation}}
\def\eeq{\end{equation}}
\newcommand{\mean}[1]{\langle #1 \rangle}
\newcommand{\const}{{\rm const}} 
\newcommand{\atanh}{\,{\rm atanh}}
\newcommand{\ie}{i.e. }
\newcommand{\eg}{e.g. }
\def\eps{\epsilon}
\def\loc{{\rm loc}} 
\def\half{\frac{1}{2}}
\def\third{\frac{1}{3}} 
\def\quarter{\frac{1}{4}}
\def\h{{\bf h}}
\def\k{{\bf k}}
\def\l{{\bf l}}
\def\m{{\bf m}}
\def\p{{\bf p}} 
\def\q{{\bf q}}
\def\r{{\bf r}}
\def\S{{\bf S}}
\def\dk{\partial_k}
\def\intinf{\int_{-\infty}^\infty}
\def\calD{{\cal D}}
\def\calO{{\cal O}}

\title{From local to critical fluctuations in lattice models: a non-perturbative renormalization-group approach}
  
\author{T. Machado} 
\author{N. Dupuis}
\affiliation{Laboratoire de Physique Th\'eorique de la Mati\`ere Condens\'ee, 
CNRS - UMR 7600, \\ Universit\'e Pierre et Marie Curie, 4 Place Jussieu, 
75252 Paris Cedex 05,  France }

\date{October 5, 2010}

\begin{abstract}
We propose a new implementation of the non-perturbative renormalization-group (NPRG) which applies to lattice models. Contrary to the usual NPRG approach where the initial condition of the RG flow is the mean-field solution, the lattice NPRG uses the (local) limit of decoupled sites as the (initial) reference system. In the long-distance limit, it is equivalent to the usual NPRG formulation and therefore yields identical results for the critical properties. We discuss both a lattice field theory defined on a $d$-dimensional hypercubic lattice and classical spin models. The simplest approximation, the local potential approximation, is sufficient to obtain the critical temperature and the magnetization of the 3D Ising, XY and Heisenberg models to an accuracy of the order of one percent. We show how the local potential approximation can be improved to include a non-zero anomalous dimension $\eta$ and discuss the Berezinskii-Kosterlitz-Thouless transition of the 2D XY model on a square lattice. 
\end{abstract}


\maketitle

\section{Introduction}

Many models of statistical physics and condensed matter are defined on a lattice. The phase diagram of a lattice model usually depends on the lattice type, the range and strength of interactions, as well as other details of the Hamiltonian. On the other hand, the precise knowledge of the Hamiltonian is often not necessary to understand the long-distance behavior of the system, in particular the universal critical properties near a second-order phase transition. In this paper, we describe an approach based on the non-perturbative renormalization group (NPRG) which captures both local and critical fluctuations in lattice models and therefore describes universal and non-universal properties. In particular, we can calculate the critical exponents, the transition temperature and the magnetization in classical spin models. 

The NPRG approach has been successfully applied to many areas of physics and in particular to the study of critical phenomena~\cite{Wetterich93,Berges02,Delamotte07}. It has recently been extended to lattice models~\cite{Dupuis08}. The strategy of the NPRG is to build a family of models indexed by a momentum scale parameter $k$, such that fluctuations are smoothly taken into account as $k$ is lowered from a microscopic scale $\Lambda$ down to 0. In practice, this is achieved by adding to the Hamiltonian (or the action) a ``regulator'' term $\Delta H_k$, which vanishes for $k=0$, and computing the corresponding Gibbs free energy (or effective action in the field theory terminology) $\Gamma_k$. The initial value $\Delta H_\Lambda$ is chosen such that in the reference system defined by the Hamiltonian $H+\Delta H_\Lambda$ all fluctuations are effectively frozen. The determination of $\Gamma_\Lambda$ then reduces to a saddle-point (mean-field) calculation. The Gibbs free energy $\Gamma_{k=0}$ we are eventually interested in is obtained from that of the reference system by solving a RG flow equation. The latter cannot in general be solved exactly (even numerically) and one has to resort to some approximations. The approximate flow equation must be sufficiently accurate (and yet tractable) to provide a good approximation of the state of the system. 

In some cases however, the mean-field solution is too far away from the actual state of the system to provide a reliable initial condition for the NPRG procedure. An example is provided by the localization transition between a Mott insulator and a superfluid in lattice boson systems. The two-pole structure of the local (on-site) propagator is crucial for the very existence of the transition. This structure is however impossible to reproduce using a RG approach starting from the mean-field (Bogoliubov) approximation. This prevents a straightforward generalization of recent NPRG studies of interacting bosons~\cite{Wetterich08,Floerchinger08,Dupuis07,Dupuis09a,Dupuis09b,Sinner09,Sinner10} to lattice models such as the Bose-Hubbard model. 

We therefore propose a new NPRG scheme for lattice models where the reference system corresponds to the (local) limit of decoupled sites. As an expansion about the local limit, the lattice NPRG is reminiscent of Kadanoff's idea of block spins~\cite{Kadanoff66}, although the way intersite interactions are progressively introduced when lowering the momentum scale $k$ makes it significantly different from a real-space RG. In the long-distance limit, the lattice NPRG is equivalent to the usual NPRG formulation and therefore yields identical results for the critical properties. 

The possibility to start from a reference system which already includes short-range fluctuations has been recognized before, and was used by Parola and Reatto in the Hierarchical Reference Theory of fluids~\cite{Parola95}, an approach which bears many similarities with the lattice NPRG. An important aspect of the lattice NPRG is that it is formulated in the field theory language commonly used in the NPRG approach. Its relation to the standard NPRG formulation is therefore obvious, and many of the approximate solutions of the flow equation satisfied by $\Gamma_k$ proposed previously also apply to the lattice case. 

In Sec.~\ref{sec_lattice} we introduce the lattice NPRG for a lattice field theory. We first recall the ``standard'' NPRG approach to lattice models~\cite{Dupuis08}, and then show that the lattice NPRG scheme merely results from a different initial condition while the long-distance (small $k$) behavior of the effective action $\Gamma_k$ remains the same. As an application, we derive a lattice field theory from the Ising model and compute the transition temperature in the local potential approximation (LPA). 
In Sec.~\ref{sec_classical_spin}, we apply the lattice NPRG to classical spin models without deriving first a lattice field theory. We find that the LPA is sufficient to obtain the critical temperature and the magnetization of the 3D Ising, XY and Heisenberg models to an accuracy of the order of 1 percent. We also discuss an improvement of the LPA (known as the LPA') which yields a non-zero anomalous dimension $\eta$. In Sec.~\ref{sec_kt}, we use the lattice NPRG to calculate the Berezinskii-Kosterlitz-Thouless (BKT)~\cite{Berezinskii70,Kosterlitz73} transition temperature of the 2D XY model on a square lattice.

\section{Lattice non-perturbative RG}
\label{sec_lattice}

We consider a lattice field theory defined on a $d$-dimensional hypercubic lattice,
\beq
H[\varphi] = \half \sum_\q \varphi_{-\q}\eps_0(\q) \varphi_\q + \sum_\r U_0(\varphi_\r), 
\label{action}
\eeq
where $\lbrace\r\rbrace$ denotes the $N$ sites of the lattice. For simplicity, we consider a one-component real field $\varphi_\r$. $\varphi_\q=N^{-1/2}\sum_\r e^{-i\q\cdot\r}\varphi_\r$ is the Fourier transformed field. The momentum $\q$ is restricted to the first Brillouin zone $]-\pi,\pi]^d$ of the reciprocal lattice. In the thermodynamic limit ($N\to\infty$), 
\beq
\frac{1}{N} \sum_\q \to \int_{-\pi}^\pi \frac{dq_1}{2\pi} \cdots \int_{-\pi}^\pi \frac{dq_d}{2\pi} \equiv \int_\q .
\eeq 
The potential $U_0$ is defined such that $\eps_0(\q=0)=0$ but is otherwise arbitrary. $\eps_0(\q)\simeq \eps_0\q^2$ for $\q\to 0$ and $\max_{\q}\eps_0(\q)=\eps_0^{\rm max}$. The lattice spacing is taken as the unit length. 

To implement the RG procedure, we add to the Hamiltonian (\ref{action}) the regulator term
\beq
\Delta H_k[\varphi] = \half \sum_\q \varphi_{-\q} R_k(\q) \varphi_\q .
\label{delta_H} 
\eeq 
Throughout the paper, we take
\beq
R_k(\q) = \bigl(\eps_k-\eps_0(\q)\bigr) \theta\bigl(\eps_k-\eps_0(\q)\bigr) 
\label{litim} 
\eeq
($\eps_k=\eps_0 k^2$), which is adapted from Ref.~\cite{Litim00} to the lattice case. The cutoff function $R_k(\q)$ leaves the high-momentum modes ($\eps_0(\q)>\eps_k$) unaffected and gives a mass $\eps_k$ to the low-energy ones (their effective (bare) dispersion satisfies $\eps_0(\q)+R_k(\q)=\eps_k$). 

In the presence of an external field, the partition function reads
\beq
Z_k[h] = \int \calD[\varphi]\, e^{-H[\varphi]-\Delta H_k[\varphi] + \sum_\r h_\r \varphi_\r} 
\eeq
and the order parameter is given by
\beq
\phi_\r = \mean{\varphi_\r} = \frac{\partial\ln Z_k[h]}{\partial h_\r} . 
\eeq
The so-called average effective action,
\beq
\Gamma_k[\phi] = - \ln Z_k[h] + \sum_\r h_\r \phi_\r - \Delta H_k[\phi] , 
\label{gamma_def} 
\eeq
is defined as a modified Legendre transform which includes the explicit subtraction of $\Delta H_k[\phi]$~\cite{Berges02}. It satisfies the exact flow equation~\cite{Wetterich93}
\beq
\partial_k \Gamma_k[\phi] = \half \sum_\q \partial_k R_k(\q) \bigl(\Gamma^{(2)}_k[\phi] + R_k\bigr)^{-1}_{\q,-\q}  
\label{flow_eq} 
\eeq
as the energy scale $\eps_k$ is varied. $\Gamma^{(2)}_k[\phi]$ is the second-order functional derivative of $\Gamma_k[\phi]$. Since $R_{k=0}(\q)=0$, $\Gamma_{k=0}[\phi]$ coincides with the effective action of the original model (\ref{action}).  

\subsection{Standard NPRG scheme}
\label{subsec_standard}

In the ``standard'' NPRG approach to lattice models~\cite{Dupuis08}, the initial value $\Lambda$ of the momentum scale $k$ is chosen such that $\eps_\Lambda$ is much larger than all characteristic energy scales of the problem. In this limit, all fluctuations are frozen and mean-field theory becomes exact: $\Gamma_{\Lambda}[\phi]=H[\phi]$. The first part of the RG procedure, when $\eps_k$ varies between $\eps_\Lambda$ and $\eps_{\kin}=\eps_0^{\rm max}$ is purely local, since the effective (bare) dispersion $\eps_0(\q)+R_k(\q)=\eps_k$ remains dispersionless for all modes. Only for $k<\kin$ does the intersite coupling start to play a role. When $k\ll 1$, \ie when $1/k$ is much larger than the lattice spacing, the lattice does not matter any more. This result is a direct consequence of the structure of the flow equation; the $\dk R_k$ term in (\ref{flow_eq}) implies that only modes with $|\q|\lesssim k$ contributes to $\dk\Gamma_k$. When $k\ll 1$, one can therefore approximate $\eps_0(\q)\simeq \eps_0\q^2$, and one recovers the flow equation of the continuum model obtained from (\ref{action}) and (\ref{delta_H}) by replacing $\eps_0(\q)$ by $\eps_0\q^2$~\cite{Dupuis08}.

\begin{figure}
\centerline{
\includegraphics[width=3.cm,clip]{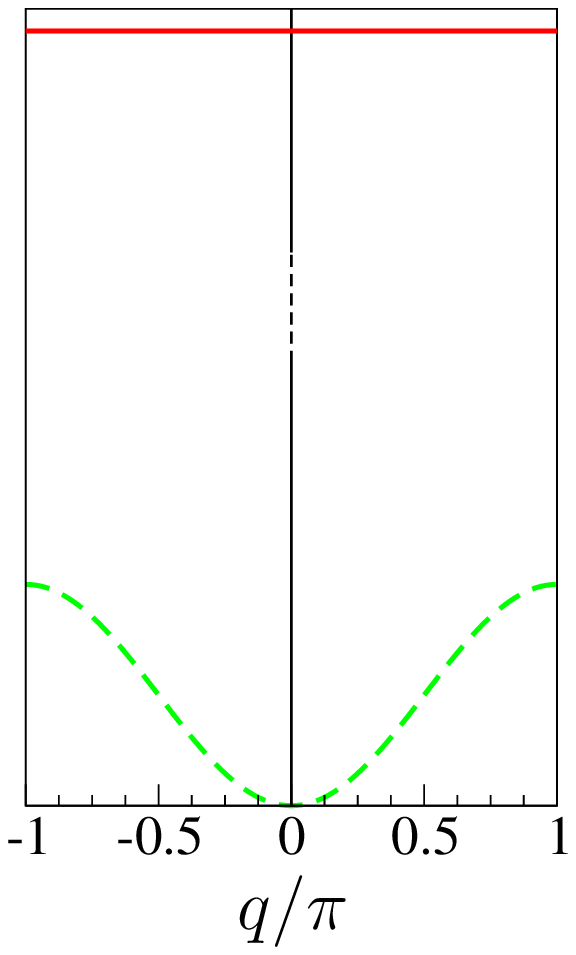}
\hspace{0.5cm}
\includegraphics[width=3.cm,clip]{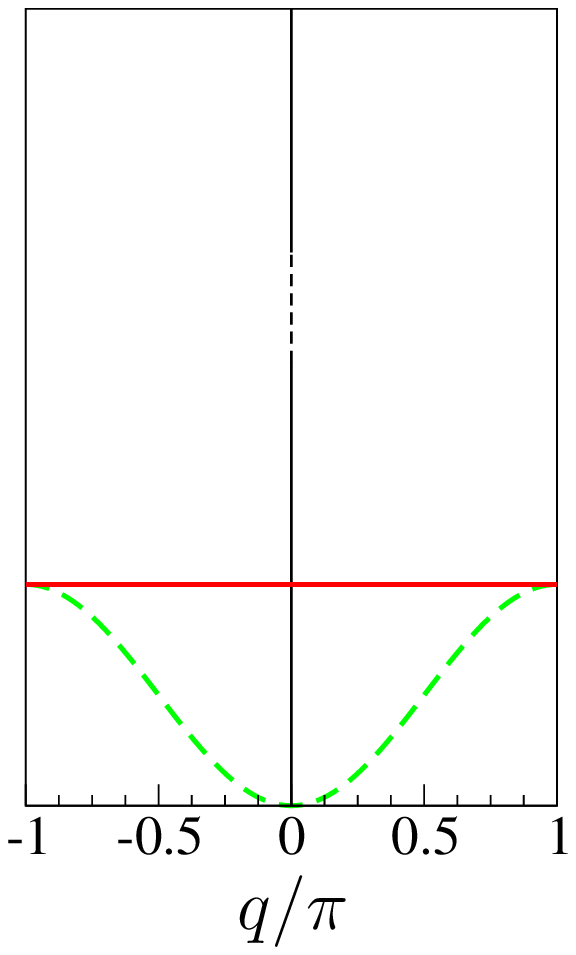}}
\caption{(Color online) Initial effective dispersion $\eps_0(q)+R_k(q)$ in the standard (left panel) and lattice (right panel) NPRG schemes ($d=1$ and $\eps_0(q)=2\eps_0(1-\cos q)$). The green dashed line shows the bare dispersion $\eps_0(q)$.}
\label{fig_dispersion_init} 
\centerline{
\includegraphics[width=2.5cm,clip]{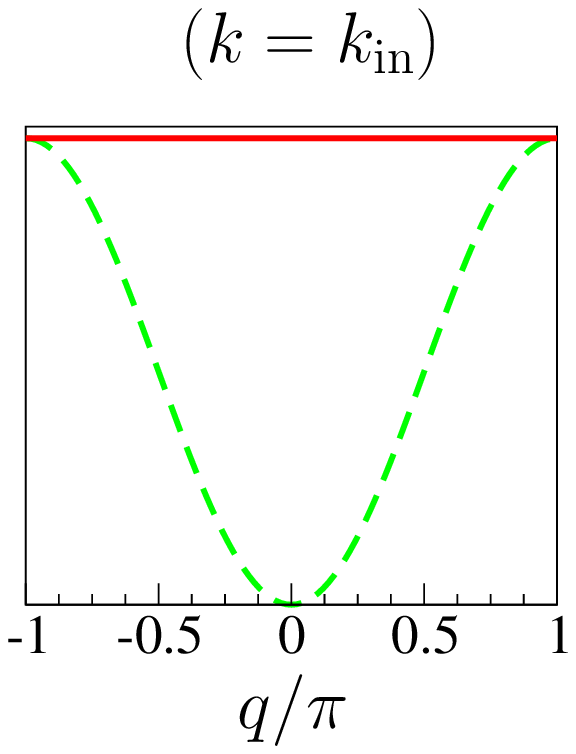}
\includegraphics[width=2.5cm,clip]{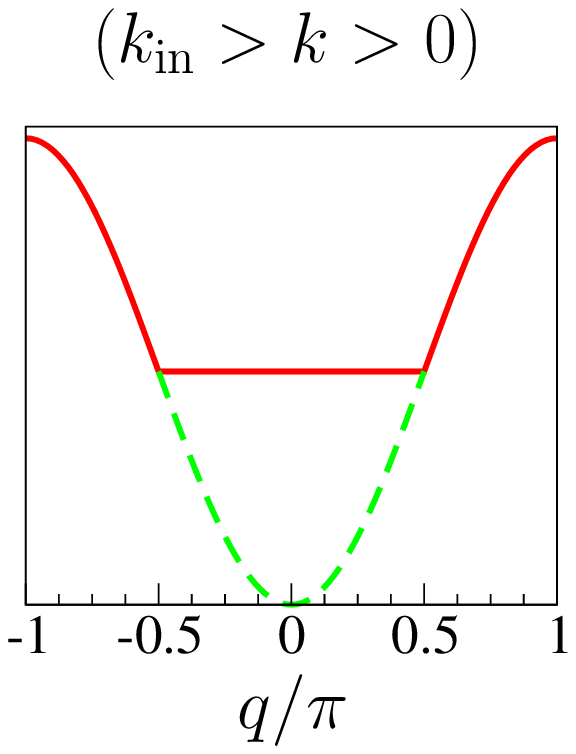}
\includegraphics[width=2.5cm,clip]{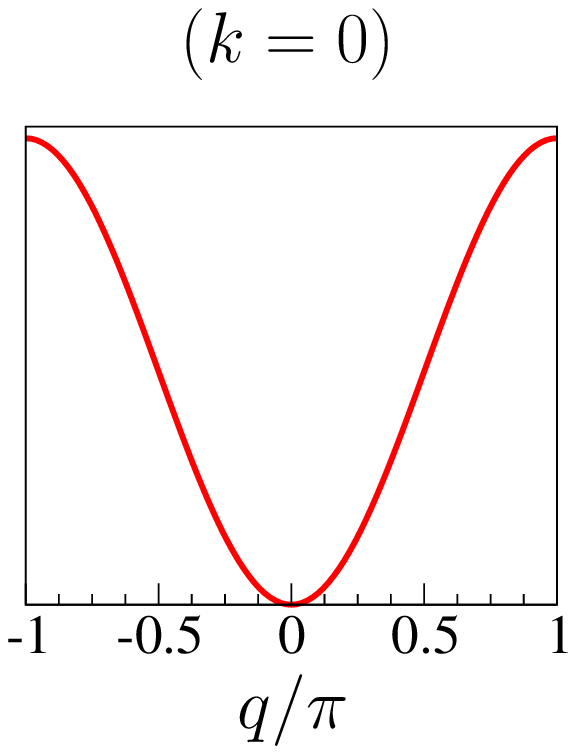}}
\caption{(Color online) Effective (bare) dispersion $\eps_0(q)+R_k(q)$ for $k=\kin$, $\kin>k>0$ and $k=0$ with the cutoff function (\ref{litim}) ($d=1$ and $\eps_0(q)=2\eps_0(1-\cos q)$). The green dashed line shows the bare dispersion $\eps_0(q)$.}
\label{fig_dispersion_eff}
\end{figure}

\subsection{Lattice NPRG scheme}
\label{subsec_lattice}  

In the lattice NPRG, we start the RG procedure from $k=\kin$, \ie we bypass the initial stage of the flow $\kin\leq k\leq \Lambda$ where the fluctuations are purely local (Fig.~\ref{fig_dispersion_init}). The average effective action $\Gamma_{\kin}[\phi]$ is no longer given by the microscopic Hamiltonian $H[\phi]$ (since the mean-field solution is not exact for the Hamiltonian $H+\Delta H_{\kin}$) but its computation reduces to a single-site problem which can be easily solved numerically (and even analytically in some models). In principle, the NPRG scheme can be defined with any cutoff function provided that the latter satisfies the initial condition~\cite{note5}
\beq
R_{\kin}(\q) = -\eps_0(\q) + C 
\label{R_init} 
\eeq
ensuring that the sites are decoupled (the limit $C\to\infty$ corresponding to the standard scheme). The choice $C=\eps_{\kin}=\eps_0^{\rm max}$ made in (\ref{litim}) is however very natural, since it allows to set up the RG procedure for $k<\kin$ in the usual way, \ie by modifying the (bare) dispersion of the low-energy modes $\eps_0(\q)<\eps_k$ without affecting the high-energy modes (Fig.~\ref{fig_dispersion_eff}). 

For $\kin>k>0$, the effective coupling in real space (defined as the Fourier transform of $\eps_0(\q)+R_k(\q)$) is long-range and oscillating. The oscillating part comes from the behavior of $R_k(\q)$ for $\eps_0(\q)\sim \eps_k$. Although the lattice NPRG is based on an expansion about the local limit, it markedly differs from Kadanoff's real-space RG~\cite{Kadanoff66} in the way degrees of freedom are progressively integrated out.  

The standard and lattice NPRG schemes thus differ only in the initial condition. Both schemes are equivalent for $k\leq\kin$ when the flow equation (\ref{flow_eq}) is solved exactly for $k>\kin$ in the standard scheme. As shown in Sec.~\ref{subsec_ising}, this is the case for classical models even within simple approximations. In practice however, one often relies on an approximate solution of the flow equation. The NPRG lattice scheme is preferable whenever the (approximate) flow equation gives a poor description of $\Gamma_{\kin}[\phi]$ starting from the mean-field result $\Gamma_{\Lambda}[\phi]=H[\phi]$. As discussed in the Introduction, this is to be expected in quantum models (such as the (Bose-)Hubbard model) where on-site (quantum) fluctuations make the local limit non-trivial. Finally we point out another advantage of the lattice NPRG; it enables to study classical spin models without first deriving a field theory (see Sec.~\ref{sec_classical_spin}).  

\subsection{Application to the Ising model} 
\label{subsec_ising} 

We consider the Ising model, 
\beq
H = - J\beta \sum_{\mean{\r,\r'}} S_\r S_{\r'} \qquad (S_\r=\pm 1) 
\label{ising_ham} 
\eeq
defined on a $d$-dimensional hypercubic lattice ($\beta=1/T$). $\mean{\r,\r'}$ denotes nearest-neighbor sites. To apply the NPRG approach, one possibility is to first derive a field theory (another, more natural, approach is described in Sec.~\ref{sec_classical_spin}). To this end, one considers the Hamiltonian
\begin{align}
H_\mu &= - J\beta \sum_{\mean{\r,\r'}} S_\r S_{\r'} - \mu\beta \sum_\r S_\r^2 \nonumber \\ 
&\equiv - \half \sum_{\r,\r'} S_\r A^{(\mu)}_{\r,\r'} S_{\r'} ,
\end{align}
which differs from that of the Ising model only by the additive constant $-\mu\beta N$. The matrix $A^{(\mu)}$ is diagonal in Fourier space with eigenvalues
\beq
\lambda_\mu(\q) = 2\beta \biggl(J \sum_{\nu=1}^d \cos q_\nu + \mu \biggr) .
\eeq
For $\mu>Jd$ the matrix $A^{(\mu)}$ is positive ($\lambda_\mu(\q)>0\;\;\forall \q$) and can be inverted. We can then rewrite the partition function of the Ising model using a Hubbard-Stratonovich transformation, 
\begin{align} 
Z_\mu &\propto \sum_{\lbrace S_\r\rbrace} \intinf \prod_\r d\varphi_\r \, e^{-\half \sum_{\r,\r'} \varphi_\r A_{\r,\r'}^{(\mu)-1} \varphi_{\r'} + \sum_\r \varphi_\r S_\r } 
\nonumber \\ 
&\propto \intinf \prod_\r d\varphi_\r \, e^{-\half \sum_{\r,\r'} \varphi_\r A_{\r,\r'}^{(\mu)-1} \varphi_{\r'} + \sum_\r \ln \cosh\varphi_\r } .
\end{align}
We thus obtain a lattice field theory with the Hamiltonian 
\begin{align}
H_\mu[\varphi] ={}& \half \sum_\q \varphi_{-\q} \left[ \frac{1}{\lambda_\mu(\q)} - \frac{1}{\lambda_\mu(0)} \right] \varphi_\q \nonumber \\ & + \sum_\r \left[ \frac{\varphi_\r^2}{2\lambda_\mu(0)} - \ln \cosh\varphi_\r \right] .
\end{align}
Rescaling the field, we can cast the Hamiltonian in the form (\ref{action}) with 
\beq
\begin{split}
\eps_0(\q) &= 2d (Jd+\mu) \frac{1-\gamma_\q}{Jd\gamma_\q+\mu} ,  \\ 
U_0(\varphi) &= \frac{Jd+\mu}{J} \varphi^2 - \ln \cosh \biggl( 2\sqrt{\frac{\beta}{J}}(Jd+\mu)\varphi \biggr) 
\end{split}
\label{app3}
\eeq
($\gamma_\q=d^{-1}\sum_\nu \cos q_\nu$). The (bare) dispersion $\eps_0(\q)$ includes long-range interactions, and $\eps_0(\pi,\pi,\cdots)=\eps_0^{\rm max}$ diverges for $\mu\to Jd$. In the limit $\mu\to\infty$ long-range interactions are suppressed and $\eps_0(\q)\to 2d(1-\gamma_\q)$. 

\begin{figure}[t]
\centerline{
\includegraphics[width=4.2cm,clip]{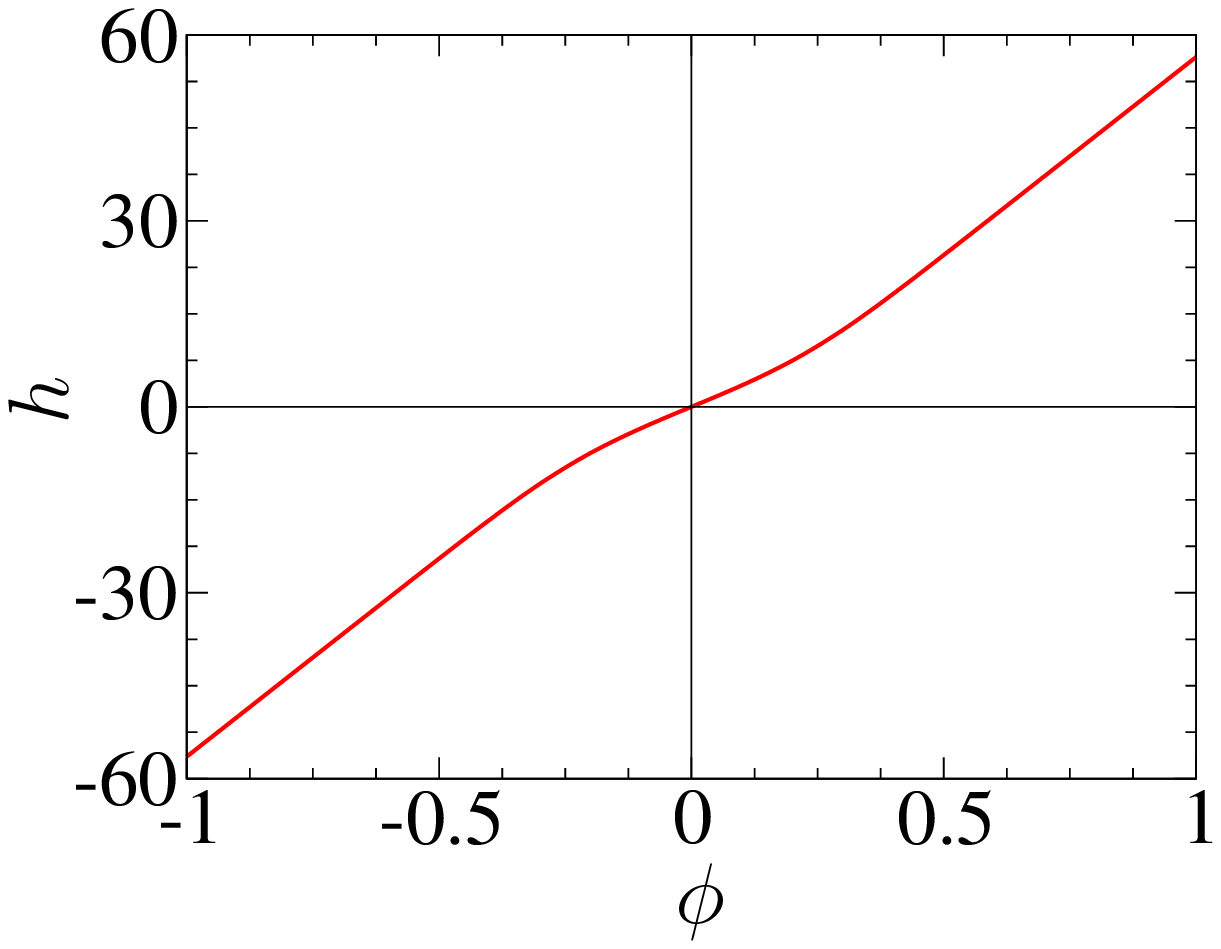}
\includegraphics[width=4.2cm,clip]{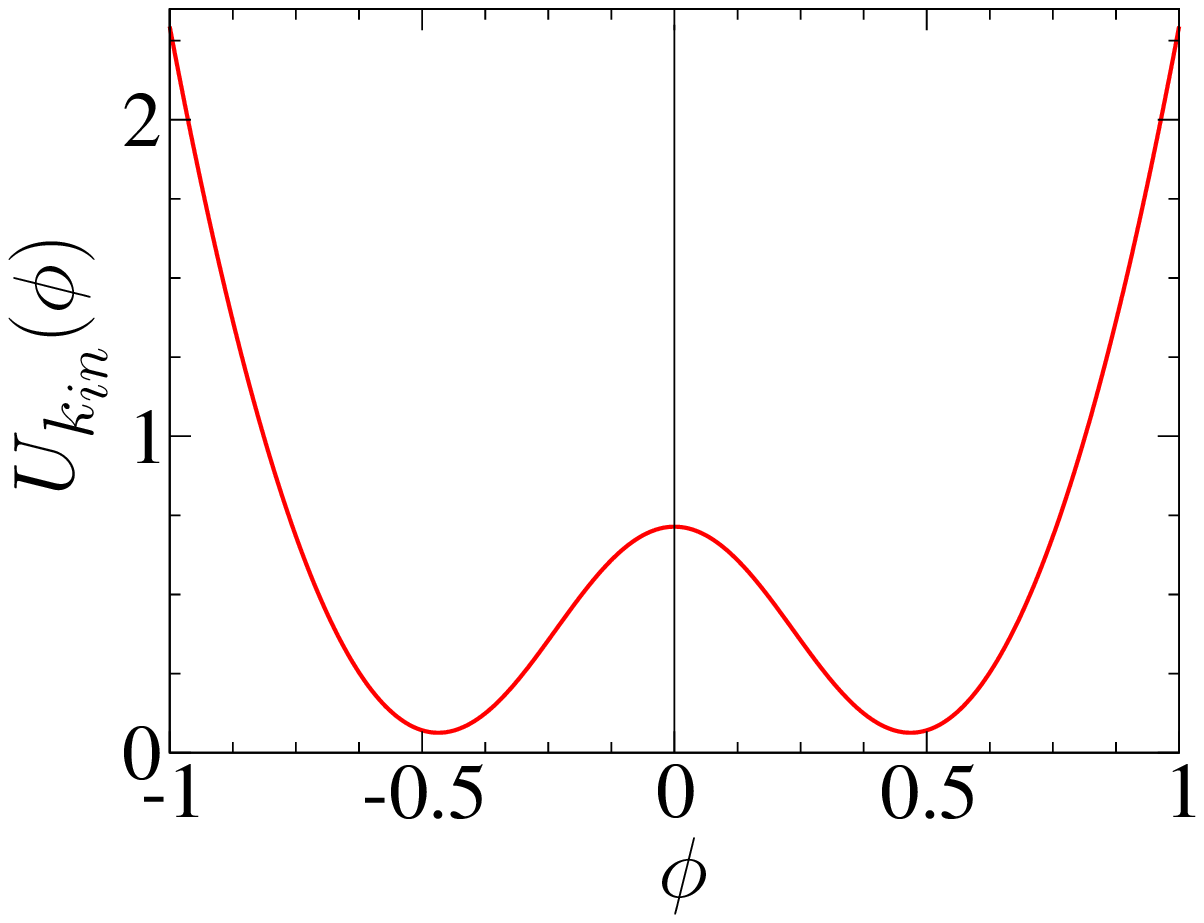}}
\caption{(Color online) Left panel: Function $h(\phi)$ obtained from the numerical solution of (\ref{phi_h}). Right panel: Effective potential $U_{\kin}(\phi)$ [Eq.~(\ref{U_kin})]. $\mu=5$, $T=4.48J$ and $d=3$.}
\label{fig_U_kin} 
\end{figure}

We are now in a position to apply the NPRG approach. In the standard scheme (Sec.~\ref{subsec_standard}), the initial value $\Gamma_\Lambda[\phi]=H[\phi]$ of the effective action is defined by the Hamiltonian (\ref{action}), with $\eps_0(\q)$ and $U_0$ given by (\ref{app3}). In the lattice NPRG scheme, the initial value $U_{\kin}$ of the effective potential has to be computed numerically. One has $Z_{\kin}[h]=\prod_\r z_{\kin}(h_\r)$, where
\beq
z_{\kin}(h) = \intinf d\varphi\, e^{- \half \eps_{\kin} \varphi^2 - U_0(\varphi) + h\varphi},
\label{lpa1} 
\eeq
is the partition function of a single site in an external field $h$. The relation between $\phi_\r$ and $h_\r$ is obtained from the equation
\beq
\phi_\r = \frac{\partial}{\partial h_\r} \ln z_{\kin}(h_\r) ,
\label{phi_h}
\eeq
which has to be computed and inverted numerically (Fig.~\ref{fig_U_kin}). The initial value of the average effective action then takes the form 
\begin{align}
\Gamma_{\kin}[\phi] &= -\sum_\r \ln z_{\kin}(h_\r) + \sum_\r h_\r\phi_\r -\Delta H_{\kin}[\phi]
\nonumber \\ &= \sum_\r U_{\kin}(\phi_\r) +  \half \sum_\q \phi_{-\q}\eps_0(\q) \phi_\q ,
\label{gam_init}
\end{align}
where 
\begin{align}
U_{\kin}(\phi) &= \frac{1}{N}\Gamma_{\kin}[\phi]\Bigl|_{\phi_\r=\phi}\nonumber \\ &= - \ln z_{\kin}(h) + h\phi - \frac{\eps_{\kin}}{2} \phi^2
\label{U_kin} 
\end{align}
is the effective potential (Fig.~\ref{fig_U_kin}). 

In the local potential approximation (LPA), one neglects the $k$-dependence of the dispersion so that
\beq
\Gamma_k[\phi] = \sum_\r U_k(\rho_\r) + \half \sum_\q \phi_{-\q}\eps_0(\q) \phi_\q .
\eeq
Here and in the following, we consider $U_k$ as a function of $\rho_\r=\phi_\r^2/2$. From (\ref{flow_eq}), one deduces
\beq
\dk U_k(\rho) = \half \int_\q \frac{\dk R_k(\q)}{\eps_0(\q) + R_k(\q) + U_k'(\rho) + 2\rho U_k''(\rho)} .
\label{U_flow_1}
\eeq
With the cutoff function (\ref{litim}), equation~(\ref{U_flow_1}) reduces to 
\beq
k \dk U_k(\rho) = \frac{\eps_k}{\eps_k+U'_k(\rho)+2\rho U''_k(\rho)} \int_\q \theta\bigl(\eps_k-\eps_0(\q)\bigr) .
\label{U_flow_2}
\eeq
The integral over $\q$ can be rewritten as
\beq
\int_\q \theta\bigl(\eps_k-\eps_0(\q)\bigr) = \int_0^{\eps_k} d\eps\, \calD(\eps) , 
\eeq
where~\cite{note2} 
\beq
\calD(\eps) = \int_\q \delta\bigl(\eps-\eps_0(\q)\bigr) .
\eeq

\begin{figure}
\centerline{
\includegraphics[height=3cm,clip]{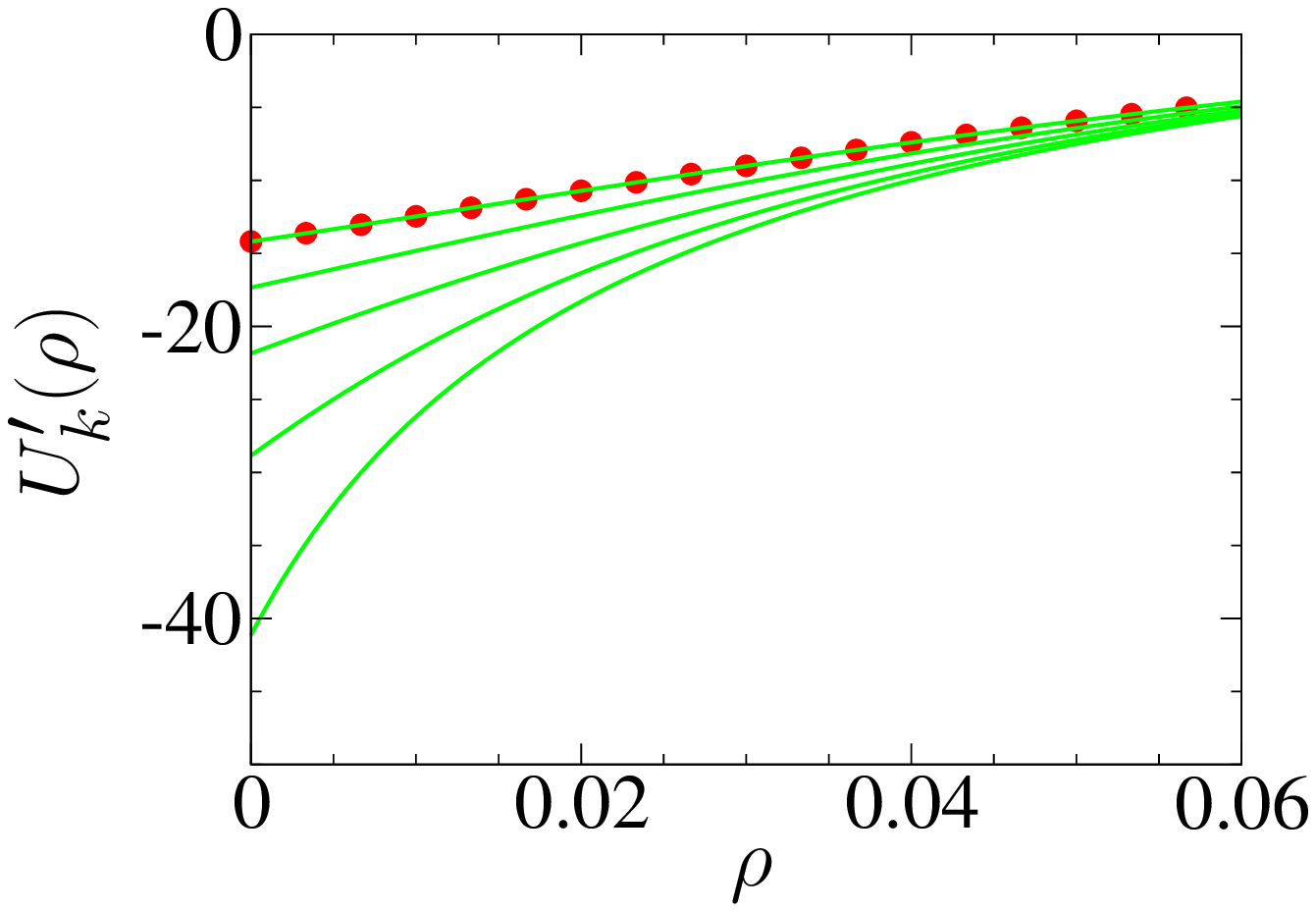}
\includegraphics[height=3cm,clip]{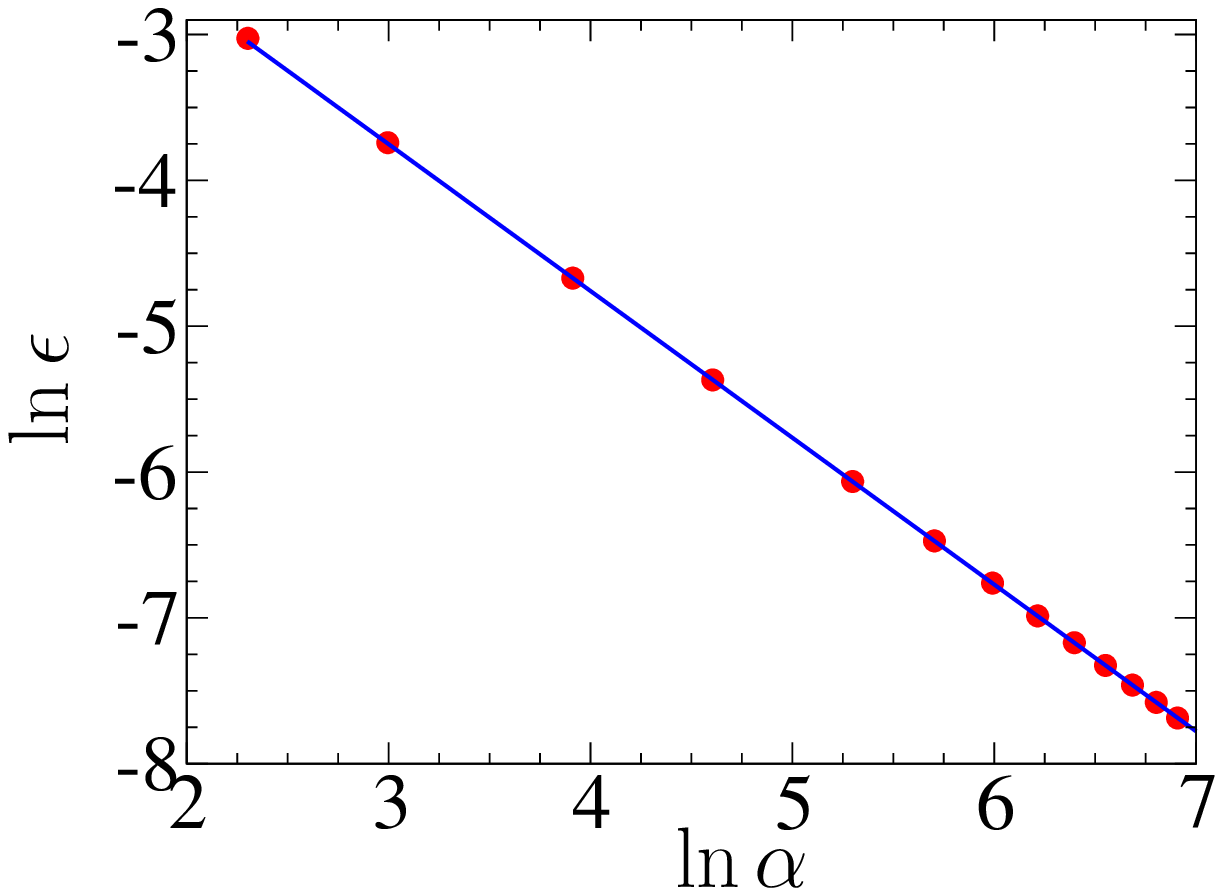}}
\caption{(Color online) Left panel: Derivative $U'_k(\rho)$ of the effective potential for various values of $k$ ranging from $\Lambda$ to $\kin$. The initial value $U_\Lambda(\rho)=U_0(\rho)$ is given by (\ref{app3}). The red points show the solution $U'_{\kin}(\rho)$ directly obtained from a numerical solution of the single-site partition function $z_{\kin}(h)$ [Eq.~(\ref{lpa1})]. Right panel: Relative error $\epsilon$ vs $\alpha=(\Lambda/\kin)^2$. $\mu=5$, $T=4.48J$ and $d=3$.}
\label{fig_lpa_local} 
\end{figure}

In the local fluctuation regime $\kin\leq k\leq\Lambda$, $H[\varphi]+\Delta H_k[\varphi]$ is a local Hamiltonian (no intersite coupling). It follows that both $-\ln Z_k[h]$ and its Legendre transform reduce to a sum of single-site contributions. The LPA for the average effective action $\Gamma_k[\phi]$ is therefore exact. We have computed the derivative $U'_k$ of the effective potential for various values of $\alpha=\eps_\Lambda/\eps_0^{\rm max}=(\Lambda/\kin)^2$. (In practice, it is easier to solve for $U'_k$ than $U_k$.) The results are shown in Fig.~\ref{fig_lpa_local} for $\alpha=1000$ (we comment on the numerical method in Sec.~\ref{subsec_spin_lpa}). As $k$ decreases from $\Lambda$ to $\kin$, the potential $U'_k(\rho)$ converges towards the exact solution obtained from the numerical computation of the local partition function $z_{\kin}(h)$ [Eqs.~(\ref{lpa1},\ref{U_kin})]. The relative error is shown in the right panel of Fig.~\ref{fig_lpa_local}. It decreases as $1/\alpha$, in agreement with the fact that the validity of the mean-field (saddle-point) approximation to $Z_\Lambda[h]$ for large $\Lambda$ is controlled by $R_\Lambda^{-1}\sim\eps_\Lambda^{-1}\sim \alpha^{-1}$. We therefore conclude that the standard and lattice NPRG schemes are equivalent in the LPA for classical lattice field theories. 

The critical temperature is obtained from the divergence of the susceptibility $\chi=1/U'_{k=0}(\rho=0)$ (or, equivalently, the divergence of the correlation length $\xi=\sqrt{\chi}$)~\cite{note7}. For $\alpha=1000$ and $d=3$ one finds that $T_c\simeq 0.747\,T_c^{\rm MF}$ is independent of $\mu$ and in very good agreement with the ``exact'' result $T_c^{\rm exact}\simeq 0.752\,T_c^{\rm MF}$ obtained from Monte Carlo simulations~\cite{Wolff89}.

\section{Classical spin models}
\label{sec_classical_spin} 

In this section, we show that the lattice NPRG can be applied to classical spin models without first deriving a field theory. For simplicity, we consider the Ising model on a $d$-dimensional hypercubic lattice [Eq.~(\ref{ising_ham})]. In the presence of an external field and a regulator term $\Delta H_k$, the partition function reads
\begin{align}
Z_k[h] &= \sum_{\lbrace S_\r\rbrace} e^{\frac{J}{T} \sum_{\mean{\r,\r'}} S_\r S_{\r'} - \half \sum_{\r,\r'} S_\r R_k(\r,\r') S_{\r'} + \sum_\r h_\r S_\r } \nonumber \\ 
&= \sum_{\lbrace S_\r\rbrace} e^{- \half \sum_\q S_{-\q} [\eps_0(\q)-2\eps_0d+R_k(\q)] S_\q + \sum_\r h_\r S_\r } ,
\label{Zising} 
\end{align}
where 
\beq
\eps_0(\q) = 2\eps_0 \sum_{\nu=1}^d (1-\cos q_\nu) , 
\eeq 
with $\eps_0=J/T$. Since $S_\r^2=1$, the term $2d\eps_0$ in (\ref{Zising}) contributes a constant term to the Hamiltonian and can be omitted. The magnetization at site $\r$ is given by 
\beq
m_\r = \mean{S_\r} = \frac{\partial \ln Z_k[h]}{\partial h_\r} 
\eeq
and the average effective action is defined by
\beq
\Gamma_k[m] = -\ln Z_k[h] + \sum_\r h_\r m_\r - \Delta H_k[m] . 
\eeq
The standard NPRG scheme cannot be used, since the partition function is not expressed as a functional integral over a continuous variable. A ``regulator'' term $\eps_k \sum_\q S_{-\q}S_\q = \eps_k \sum_\r S_\r^2=N\eps_k$ would only add a constant term to the Hamiltonian. On the contrary, there is no difficulty to apply the lattice NPRG scheme. With the cutoff function (\ref{litim}), one has
\beq
Z_{\kin}[h] = \sum_{\lbrace S_\r\rbrace} e^{-2d\eps_0 N + \sum_\r h_\r S_\r} = e^{-2d\eps_0 N} \prod_\r z(h_\r) , 
\eeq
where
\beq 
z(h) = \sum_{S=\pm 1} e^{hS} = 2 \cosh(h) 
\eeq
is the partition function of a single site in an external field $h$. The magnetization at site $\r$,
\beq
m_\r = \frac{\partial}{\partial h_\r} \ln z(h_\r) = \tanh(h_\r) ,
\eeq
varies between -1 and 1. Up to an additive constant, we obtain
\beq
\Gamma_{\kin}[m] = \sum_\r U_{\kin}(\rho_\r) + \half \sum_\q m_{-\q} \eps_0(\q) m_\q 
\label{gam_init_1}
\eeq
and the effective potential~\cite{note3}  
\beq
U_{\kin}(\rho) = \half \ln(1-2\rho) + \sqrt{2\rho} \atanh(\sqrt{2\rho}) -4d\eps_0\rho ,
\eeq
where $\rho_\r=m_\r^2/2$. $U_{\kin}(1/2)=\ln(2)-2d\eps_0$ is finite but $U'_{\kin}(\rho) \sim -\half \ln(1-2\rho)$ diverges for $\rho\to 1/2$. This divergence suppresses the propagator
\beq
\frac{1}{\eps_k+~R_k(\q)+U'_k(\rho)+2\rho U''_k(\rho)} 
\eeq
appearing in the flow equation (\ref{U_flow_1}) and therefore the fluctuations corresponding to a large magnetization.

A comment is in order here. We have followed the usual convention to define the average effective action $\Gamma_k$ as a modified Legendre transform which includes the explicit subtraction of $\Delta H_k[m]$ [Eq.~(\ref{gamma_def})]~\cite{Berges02}. The definition of the average effective action $\Gamma_k$ is of course arbitrary provided that $\Gamma_{k=0}$ corresponds to the true Legendre transform of the original model. Since
\beq
U_{\kin}(\rho) = \rho(1-4d\eps_0) + \calO(\rho^2) 
\eeq
for $\rho\to 0$, we find that the initial transition temperature is determined by $1=4d\eps_0$, \ie $T_c^{(\kin)}=4dJ$, which differs from the mean-field transition temperature $T_c^{\rm MF}=2dJ$ by a factor of 2.  $\Gamma_{\kin}[m]$ assumes a mean-field treatment of the intersite coupling term $\half \sum_\q S_{-\q} \eps_0(\q) S_\q$ as in the usual mean-field approach to the Ising model. The discrepancy between $T_c^{(\kin)}$ and $T_c^{\rm MF}$ comes from the fact that $\half \sum_\q S_{-\q} \eps_0(\q) S_\q$ includes a local term $\half \sum_\r 2d\eps_0 S_\r^2=Nd\eps_0$. The latter contributes a mere constant to the Hamiltonian, but is considered at the mean-field level in the average effective action where it gives a term $-d\eps_0\sum_\r m_\r^2$. To make contact with the usual mean-field theory, we consider the effective average action 
\beq
\bar\Gamma_k[m] = \Gamma_k[m] + \half \sum_{\r} R_k(\r,\r) m_\r^2 
\label{gamma_bar} 
\eeq
and the corresponding effective potential 
\beq
\bar U_k(\rho) = U_k(\rho) + \rho R_k(\r,\r) . 
\label{U_bar_def} 
\eeq
$\bar\Gamma_k[m]$ differs from the true Legendre transform only by non-local terms. The initial value 
\beq
\bar U_{\kin}(\rho)=\rho(1-2d\eps_0)+\calO(\rho^2)
\eeq
reproduces the mean-field result $T_c^{(\kin)}=T_c^{\rm MF}$. Again we stress that $\Gamma_k$ and $\bar\Gamma_k$ lead to the same $k=0$ results and in particular to the same critical temperature.

\begin{table}[t]
\renewcommand{\arraystretch}{1.5}
\begin{center}
\begin{tabular}{|c||c|c|c|}
\hline 
& $T_c^{\rm MF}$ & $T_c^{\rm exact}$ & $T_c^{\rm NPRG}$ 
\\ \hline 
Ising 3D & 6 & 4.51 & 4.48 
\\ \hline 
XY 3D & 3 & 2.20 & 2.18 
\\ \hline 
Heisenberg 3D & 2 & 1.44 & 1.42 
\\ \hline  
\end{tabular}
\end{center}
\caption{Critical temperature $T_c^{\rm NPRG}$ obtained in the LPA compared to the mean-field estimate $T_c^{\rm MF}$ and the Monte-Carlo result $T_c^{\rm exact}$~\cite{Wolff89,Hasenbush90,Holm93}. All temperatures are in unit of $J$.}
\label{table_Tc}
\end{table}

\begin{figure}
\centerline{\includegraphics[width=6cm,clip]{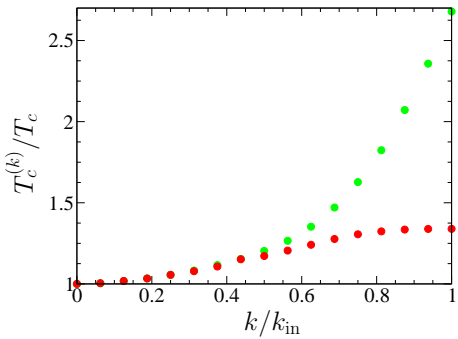}}
\caption{(Color online) Transition temperature $T_c^{(k)}$ obtained from the effective potential $U_k$ (green points) and $\bar U_k$ (red points).}
\label{fig_Tc_k} 
\end{figure}

\begin{figure}[t]
\centerline{
\includegraphics[height=3.cm,clip]{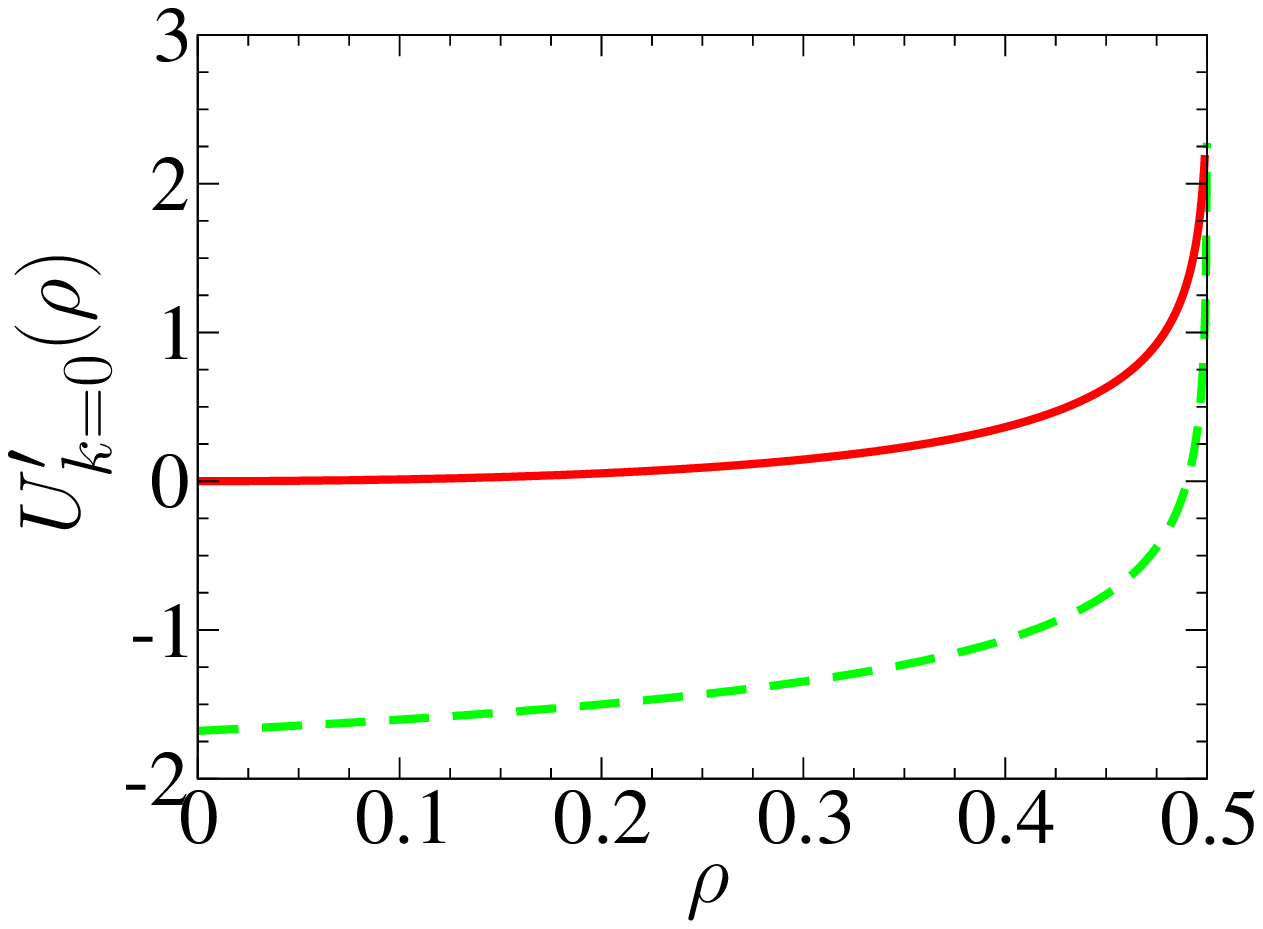}
\includegraphics[height=3.cm,clip]{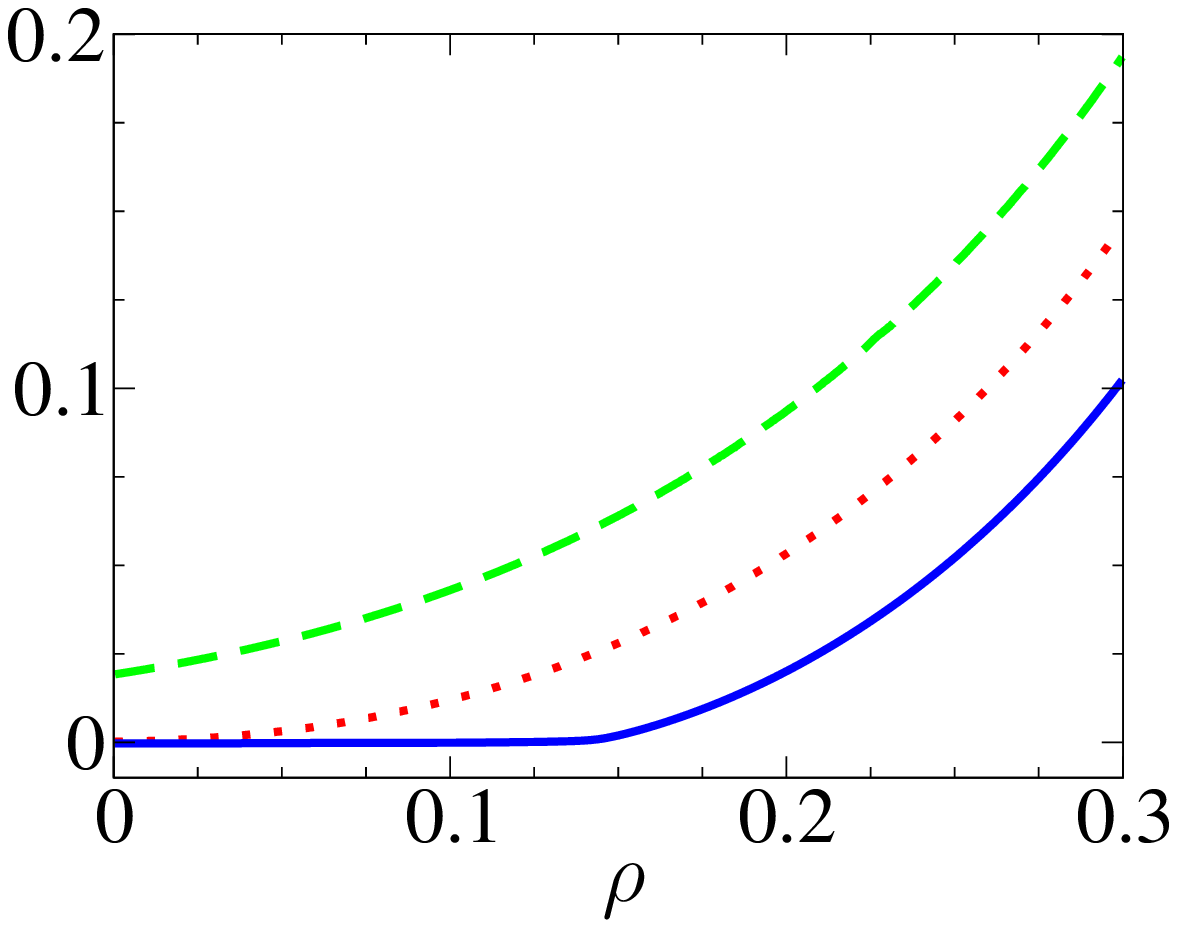}}
\caption{(Color online) Left panel: Potentials $U'_{k=0}(\rho)$ (red solid line) and $U'_{\kin}(\rho)$ (green dashed line) at criticality ($T=T_c$) in the LPA ($d=3$). Right panel: Potential $U'_{k=0}(\rho)$ for $T=T_c$ (red dotted line), $T=1.05\,T_c$ (green dashed line) and $T=0.95\,T_c$ (blue solid line).} 
\label{fig_U_1} 
\end{figure}

\subsection{Local Potential Approximation}
\label{subsec_spin_lpa} 

We have solved the equation for $U'_k(\rho)$ (see Eq.~(\ref{U_flow_2})) numerically using Euler's method with a typical RG time step $\Delta t=-10^{-4}$ ($t=\ln(k/\kin)$). The function $U'_k(\rho)$ is discretized with a few hundreds points in $\rho$. The convexity of the potential (see Fig.~\ref{fig_U_1}) makes the numerical resolution difficult below $T_c$ (in particular at low temperatures) and there is a tendency to numerical instability for
large values of $|t|$. However, the value $\rho_0$ for which $U'_k(\rho)=0$ usually converges before instability problems arise.

The critical temperature is obtained from the criterion $U_{k=0}'(\rho=0)=0$ (Sec.~\ref{subsec_ising}). One finds $T_c \simeq 0.747 \,T_c^{\rm MF}$ for the three-dimensional Ising model, in very good agreement with the results of Sec.~\ref{subsec_ising} and the ``exact'' result $T_c^{\rm exact}\simeq\,0.752\,T_c^{\rm MF}$ obtained from Monte-Carlo simulations~\cite{Wolff89}. We have obtained a similar accuracy for the critical temperature of the XY and Heisenberg models in $d=3$ (Table~\ref{table_Tc}).

Fig.~\ref{fig_Tc_k} shows the transition temperature $T_c^{(k)}$ of the three-dimensional Ising model deduced from the effective potentials $U_k$ and $\bar U_k$ [Eq.~(\ref{U_bar_def})]. As $k$ decreases, $T_c^{(k)}$ converges rapidly towards the actual transition temperature $T_c=T_c^{(k=0)}$. This result is due to the fact that all degrees of freedom contribute more or less equally to the thermodynamics. Once $k\ll \kin$, thermodynamic quantities are therefore obtained with a reasonable accuracy. This also explains why the LPA, which does not correctly describe the long-distance limit of the propagator when $T\simeq T_c$, is remarkably successful in computing the transition temperature and other thermodynamic quantities. 

Fig.~\ref{fig_U_1} shows the derivative $U'_k(\rho)$ of the effective potential for $k=\kin$ and $k=0$ at criticality ($T=0.747 \,T_c^{\rm MF}$), as well as $U'_{k=0}(\rho)$ for $T=T_c$, $T>T_c$ and $T<T_c$. In the latter case, we find $U'_{k=0}(\rho)=0$ and therefore $U_{k=0}(\rho)=\const$ for $\rho\leq \rho_0$, where $\rho_0=m_0^2/2$ determines the actual magnetization $m_0$ of the system. This result is a consequence of the convexity of the potential in the low-temperature phase, a property which is known to be satisfied in the LPA~\cite{Berges02}. 

\begin{figure}
\centerline{\includegraphics[width=6cm,clip]{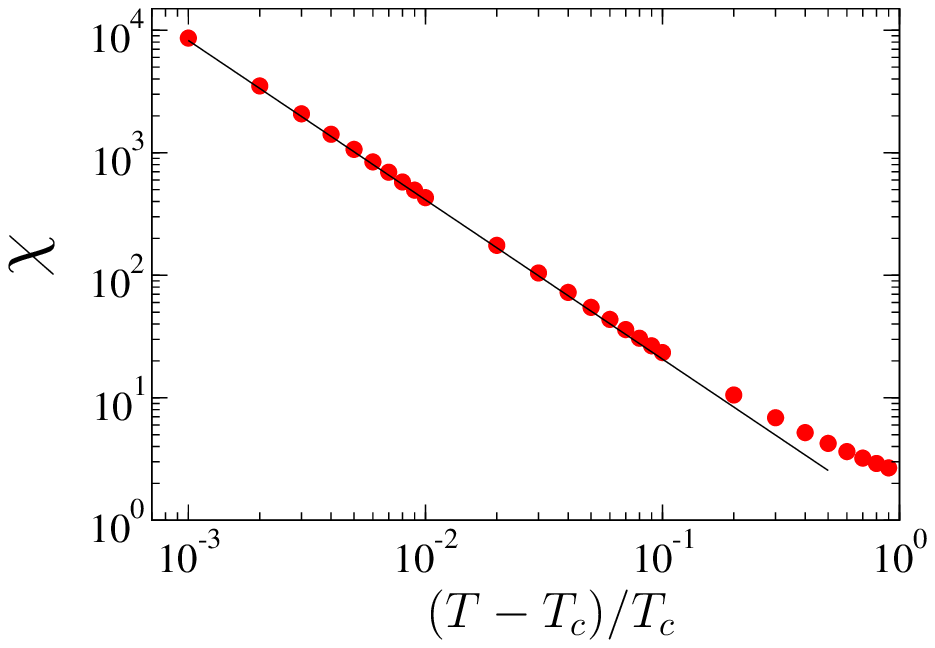}}
\vspace{0.3cm}
\centerline{\includegraphics[width=5.5cm,clip]{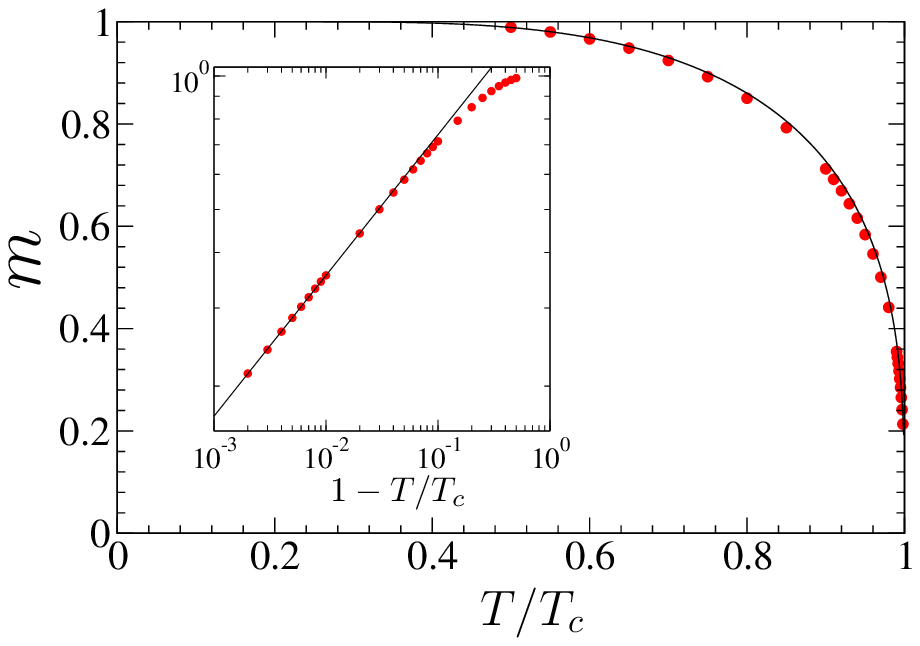}}
\caption{(Color online) Top panel: Uniform susceptibility $\chi$ (red points) in the high-temperature phase and a fit $\chi\propto (T-T_c)^{-\gamma}$ with $\gamma=2\nu\simeq 1.30$. Bottom panel: Magnetization $m$ in the low-temperature phase (red points) and the Essam-Fisher approximant~\cite{Essam63} (solid line). The inset shows a fit to $m\propto(T_c-T)^\beta$ with $\beta=\nu/2\simeq 0.32$.}
\label{fig_chi_m} 
\end{figure}

In Fig.~\ref{fig_chi_m}, we show the uniform susceptibility $\chi=1/U'_{k=0}(\rho=0)$ in the high-temperature phase~\cite{note7}, as well as the magnetization below $T_c$ with the Essam-Fisher approximant~\cite{Essam63}. We find the critical exponents $\nu=2\beta=\gamma/2\simeq 0.64-0.65$ (with $\eta=0$ in the LPA), in agreement with the known result in the LPA with the cutoff function (\ref{litim})~\cite{Litim00,Bervillier08}.

Not surprisingly, the LPA is not as accurate in two dimensions. For the 2D Ising model, we find $T_c\simeq 0.48\,T_c^{\rm MF}$, to be compared with the exact result $T_c^{\rm exact}=2J/\ln(1+\sqrt{2})
\simeq 0.567\,T_c^{\rm MF}$~\cite{Onsager44}.

\subsection{Renormalization of the spectrum}
\label{subsec_A_k} 

A natural generalization of the LPA includes a renormalization of the amplitude of the spectrum. We therefore consider the Ansatz
\beq
\Gamma_k[m] = \sum_\r U_k(\rho_\r) + \half \sum_\q A_k \eps_0(\q) m_{-\q} m_\q .
\label{gam_spec}
\eeq
This approximation can be seen as the first step of a circular harmonic expansion of the renormalized dispersion $\eps(\q)$~\cite{Dupuis08}. Since 
\beq
\Gamma^{(2)}_k(\q;\rho) = A_k \eps_0(\q) + U'_k(\rho) + 2\rho U''_k(\rho)
\eeq
in a uniform field $\phi_\r=\sqrt{2\rho}$, we can define the renormalized spectrum amplitude by
\beq
A_k = \frac{1}{\eps_0} \Gamma_k^{(2)}(\r-\r';\rho_{0,k})
\label{A_def} 
\eeq
where $\r$ and $\r'$ are nearest neighbors. The amplitude $A_k\equiv A_k(\rho_{0,k})$ should be understood as the first term in the expansion of the function
\beq
A_k(\rho) = A_k(\rho_{0,k}) + A_k^{(1)}(\rho_{0,k}) (\rho-\rho_{0,k}) + \cdots 
\eeq
about the minimum $\rho_{0,k}$ of the effective potential $U_k(\rho)$. Another possible definition of the spectrum amplitude is $A_k\equiv A_k(\bar\rho_{0,k})$, where $\bar\rho_{0,k}$ is the minimum of $\bar U_k(\rho)$ [Eq.~(\ref{U_bar_def})]. The flow equation for $A_k$ follows from (\ref{flow_eq}) and (\ref{A_def}),
\begin{align}
\dk A_k ={}& \frac{\gamma_3^2}{\eps_0} \int_\q \left(1-\frac{\eps_0(\q)}{2d\eps_0}\right) \dk R_k(\q) G(\q)^2 \nonumber \\ &\times \int_\p \left(1-\frac{\eps_0(\p)}{2d\eps_0}\right) G(\p) ,
\label{Ak_flow} 
\end{align} 
where $\gamma_3=\sqrt{2\rho_{0,k}}[3U_k''(\rho_{0,k})+2\rho_{0,k}U_k'''(\rho_{0,k})]$. The flow equation for $U_k(\rho)$ is identical to (\ref{U_flow_1}) with $\eps_0(\q)$ replaced by $A_k\eps_0(\q)$. 

\begin{figure}[t]
\centerline{
\includegraphics[height=2.9cm,clip]{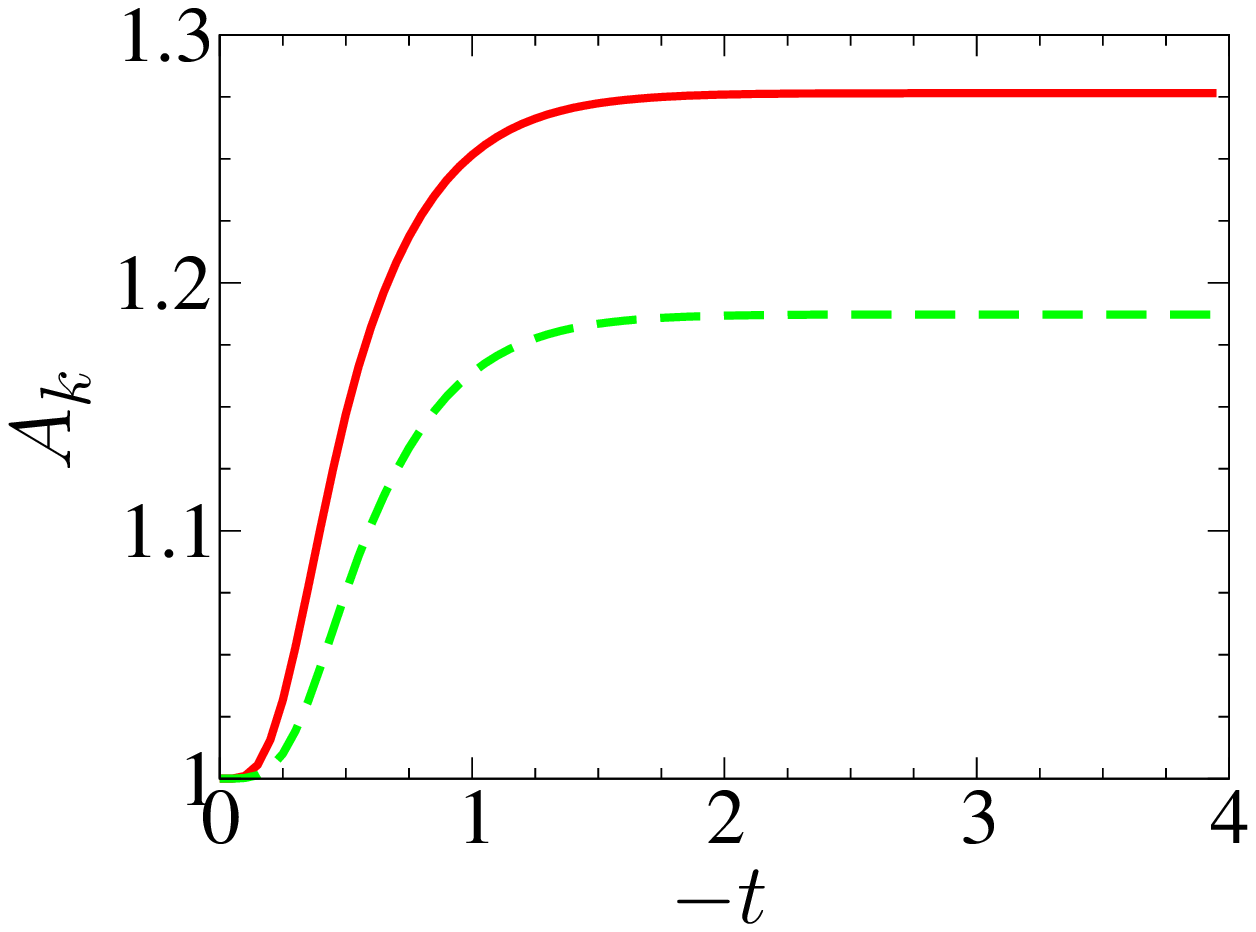}
\hspace{0.1cm}
\includegraphics[height=2.9cm,clip]{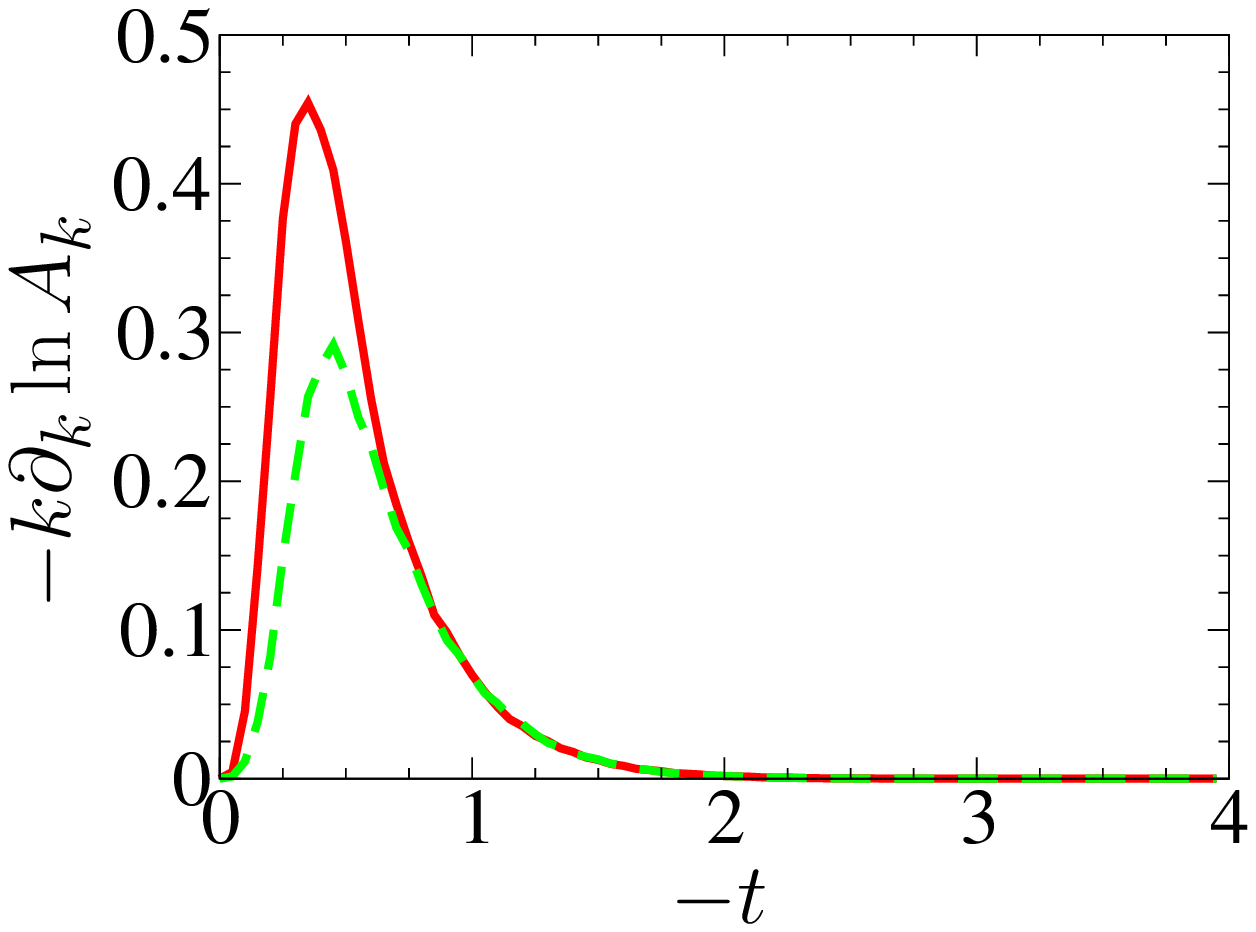}}
\caption{(Color online) Left panel: $A_k(\rho_{0,k})$ (red solid line) and $A_k(\bar\rho_{0,k})$ (green dashed line) vs $-t=\ln(\kin/k)$. Right panel: $-k\dk \ln A_k(\rho_{0,k})$ (red solid line) and $-k\dk \ln A_k(\bar\rho_{0,k})$ (green dashed line) vs $-t$. All curves are obtained at criticality ($T=T_c$) and for $d=3$.}
\label{fig_A_k}  
\end{figure}

Numerical results for $A_k(\rho_{0,k})$ and $A_k(\bar\rho_{0,k})$ are shown in Fig.~\ref{fig_A_k}. In both cases, $A_k$ varies when $k\sim 1$, in agreement with the expectation that the amplitude of the harmonic $\cos(nq_\nu)$ should vary when $k\sim 1/n$~\cite{Dupuis08}. The variation of $A_k$ is moderate (from 1 to 1.27 for $A_k(\rho_{0,k})$ and from 1 to 1.19 for $A_k(\bar\rho_{0,k})$) and weakly affects the critical temperature which remains within a few percents of the exact result: $T_c=0.74 \,T_c^{\rm MF}$ with $A_k(\rho_{0,k})$ and $T_c=0.72\, T_c^{\rm MF}$ with $A_k(\bar\rho_{0,k})$. We expect that the inclusion of additional higher-order harmonics in the spectrum would give a better estimate of $T_c$.

\subsection{LPA'}
\label{subsec_lpap} 

When $k\ll 1$, we can approximate $\eps_0(\q)$ by $\eps_0\q^2$. In this regime, a simple improvement over the LPA (known as the LPA') consists in including a field renormalization factor $Z_k$ so that the renormalized dispersion $\eps(\q)$ is given by $Z_k\eps_0(\q)\simeq Z_k\eps_0\q^2$. The LPA' can be generalized to all values of $k$ by writing the average effective action as~\cite{Dupuis08}
\beq
\Gamma_k[m] = \sum_\r U_k(\rho_\r) + \half \sum_\q m_{-\q} Z_k \eps_0(\q) m_\q . 
\eeq
Although this Ansatz is formally similar to (\ref{gam_spec}), $Z_k$ should not be confused with the amplitude $A_k$ introduced in the preceding section. $Z_k$ is computed from the $\calO(\q^2)$ part of the spectrum,
\beq
Z_k = \frac{1}{\eps_0} \lim_{\q \to 0} \frac{\partial}{\partial \q^2} \Gamma^{(2)}_k(\q;\rho_{0,k}) ,
\label{Z_def}  
\eeq
and therefore receives contributions from all harmonics. The LPA' can be justified when $k\sim 1$ by noting that in this limit the renormalization of the spectrum is weak ($Z_k\sim 1$), so that the approximation $\eps(\q)\simeq Z_k\eps_0(\q)$, valid for small $\q$, is expected to remain approximately valid in the whole Brillouin zone~\cite{Dupuis08}. Nevertheless, because short-range fluctuations are important for the thermodynamics (Sec.~\ref{subsec_spin_lpa}), the LPA' might lead to a slight deterioration of the value of $T_c$ obtained in the LPA. As in the preceding section, we can compute $Z_k$ either from the minimum of $U_k(\rho)$ (as in (\ref{Z_def})) or from the minimum $\bar\rho_{0,k}$ of $\bar U_k(\rho)$. 

\begin{figure}
\centerline{\includegraphics[width=5.cm,clip]{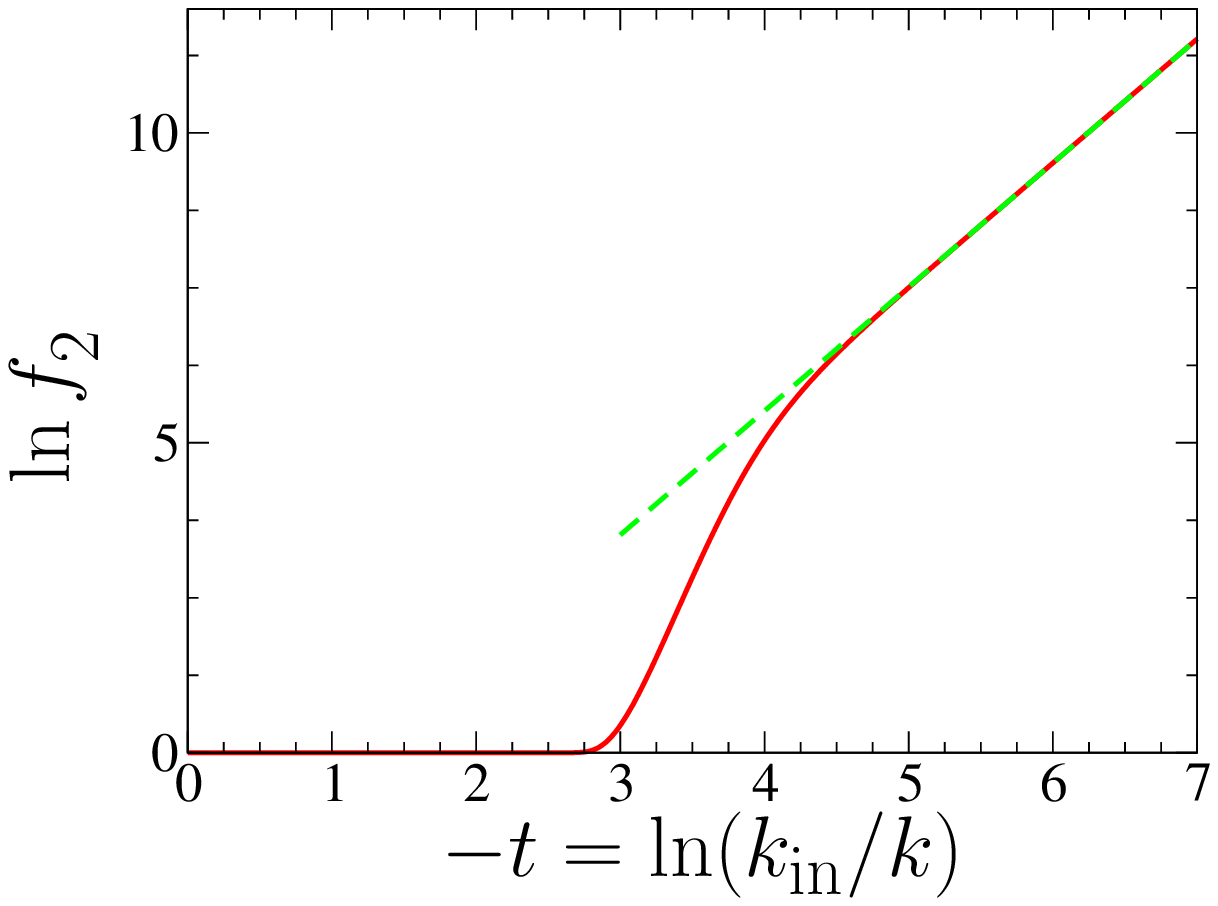}}
\caption{(Color online) The function $f_2(k)=k^{-d}g_2(k)$ vs $-t=-\ln(k/\kin)$ for $n=6$ and $t_c=-3$ ($d=2$). The green dashed line corresponds to $g_2(k)=1$.}
\label{fig_g2} 
\includegraphics[width=5.5cm,clip]{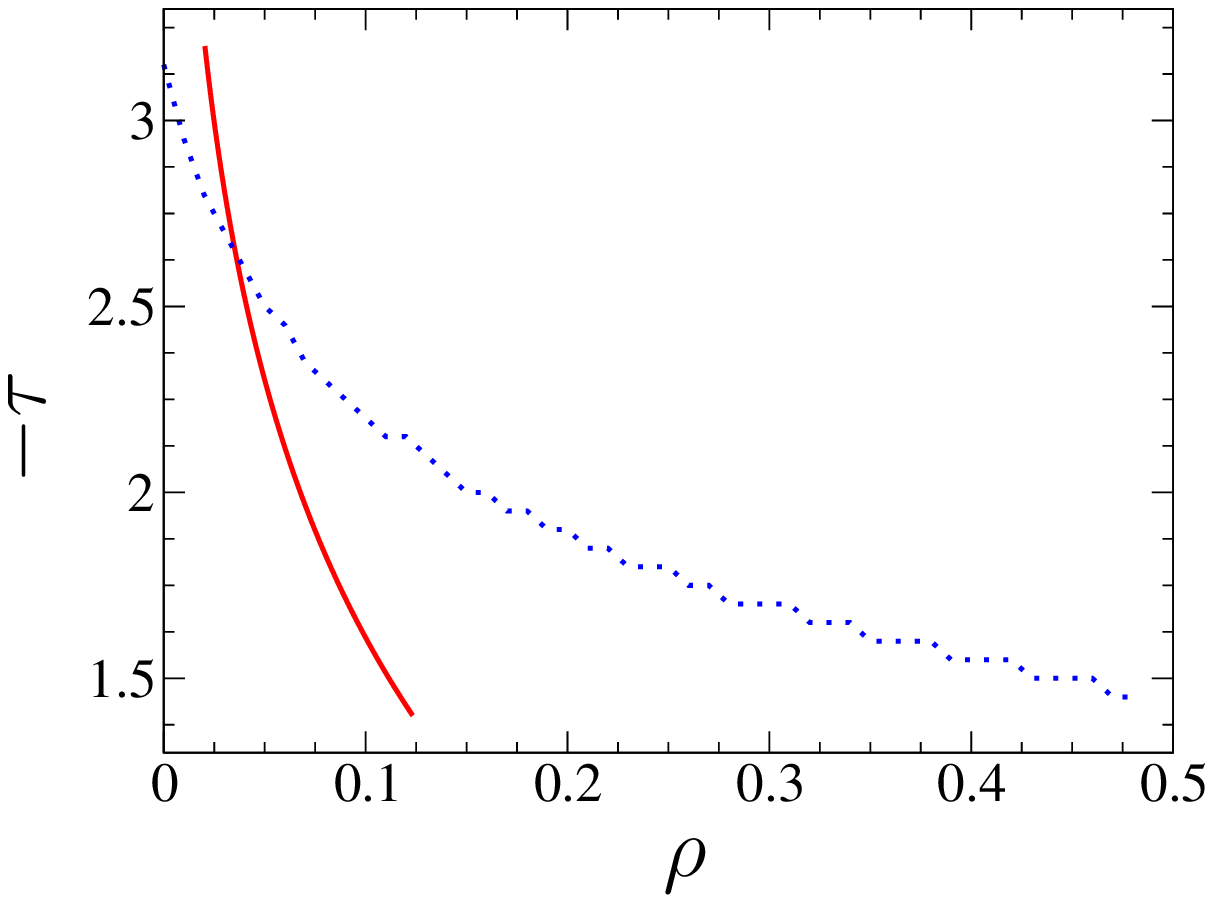}
\caption{(Color online) RG time $-\tau(\rho)$ beyond which the potential $U'_k(\rho)$ differs from $U'_{k=0}(\rho)$ by less then $10^{-3}$ (blue dotted line). The red solid line shows $\rho=\half e^{\tau}$ (see text).}
\label{fig_taurho} 
\end{figure}

To obtain a fixed point when the system is critical, one should redefine the cutoff function, 
\beq
R_k(\q) = Z_k \bigl(\eps_k-\eps_0(\q)\bigr) \theta\bigl(\eps_k-\eps_0(\q)\bigr) ,
\eeq
and introduce the dimensionless variables
\beq 
\tilde\rho = Z_k k^{-d} \eps_k \rho, \qquad \tilde U_k(\tilde\rho) = k^{-d} U_k(\rho) . 
\label{tilde_def} 
\eeq
This change of variables cannot be done at the beginning of the flow if one works with a given range of $\tilde\rho$ values. This would indeed correspond to a smaller and smaller range in $\rho$, whereas a good determination of the critical temperature requires to consider the window $0\leq\rho \leq 1/2$. To circumvent this difficulty, we define
\beq
\tilde\rho = Z_k k^{-d} \eps_k g_1(k) \rho, \qquad \tilde U_k(\tilde\rho) = k^{-d} g_2(k) U_k(\rho) ,
\label{tilde_def_new}  
\eeq
where
\beq
\begin{split}
g_1(k) &= \left\lbrace \begin{array}{lll}  k^d (Z_k\eps_k)^{-1} & \mbox{if} & k \gg k_c , \\ 
                       1 & \mbox{if} & k \ll k_c , 
                       \end{array} \right. \\ 
g_2(k) &= \left\lbrace \begin{array}{lll} k^d & \mbox{if} & k \gg k_c , \\ 
                       1 & \mbox{if} & k \ll k_c .
                       \end{array} \right. 
\end{split}
\eeq
For $k\gg k_c$, $\tilde\rho$ and $\tilde U_k$ are equal to $\rho$ and $U_k$, whereas they coincide with the dimensionless variables (\ref{tilde_def}) when $k\ll k_c$. The momentum scale $k_c$ will be determined below. In practice, we take
\beq
\begin{split}
g_1(k) &= [x+k^{-d} Z_k\eps_k(1-x)]^{-1} , \\ 
g_2(k) &= [x+k^{-d} (1-x)]^{-1} , 
\end{split}
\eeq
where $x=e^{-(k/k_c)^n}$ with $n>d$. The parameter $n$ fixes the size of the crossover region $k\sim k_c$ between dimensionful and dimensionless variables (Fig.~\ref{fig_g2}). 

Fig.~\ref{fig_taurho} shows the RG time $-\tau(\rho)$ beyond which the potential $U'_k(\rho)$ differs from $U'_{k=0}(\rho)$ by less than $10^{-3}$ ($d=3$). (The RG time $t$ is defined by $k=\kin e^t$.) This precision ($10^{-3}$) is sufficient to determine the critical temperature to an accuracy of the order of 1 percent. As expected, the potential converges to its asymptotic value faster for large values of $\rho$. In Fig.~\ref{fig_taurho}, the red solid line shows the maximum value $\rho_{\rm max}=\half e^{\tau}$ at time $\tau$ if one works with a fixed window $\tilde\rho\in [0,1/2]$ [Eq.~(\ref{tilde_def})] (we neglect $Z_{k=e^\tau}\simeq 1$ in the calculation of $\rho_{\rm max}$). Thus we see that a natural choice for $t_c=\ln(k_c/\kin)$ is $t_c\simeq -2.5$, since for times $-t>-t_c$ the window $[0,\rho_{\rm max}]$ becomes larger than the range of $\rho$ values for which the potential $U'_k(\rho)$ has not converged to its asymptotic value yet. In practice, we verify that our results are independent of the precise choice of $k_c$ and $n$. 

\begin{figure}
\centerline{
\includegraphics[width=4cm,clip]{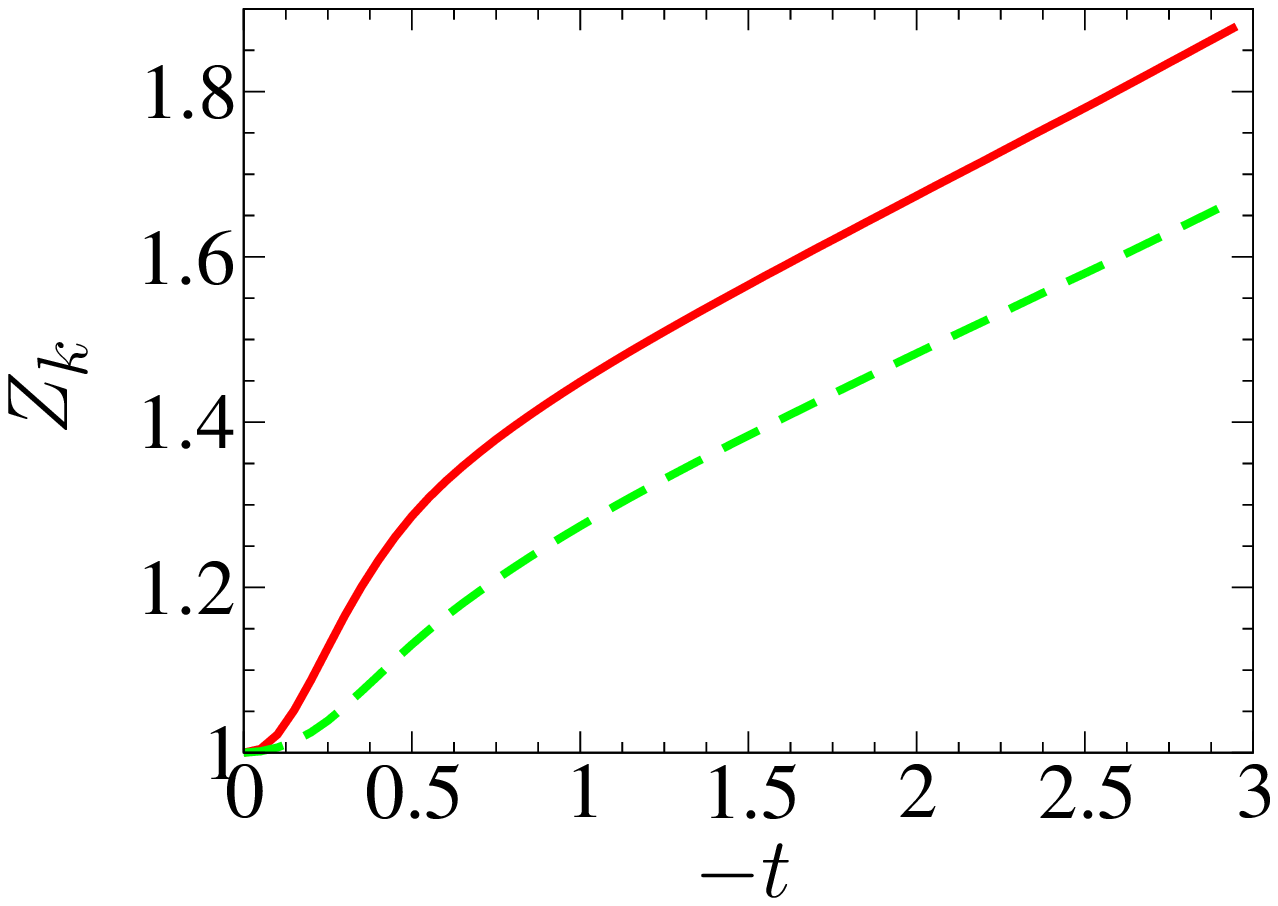}
\includegraphics[width=4cm,clip]{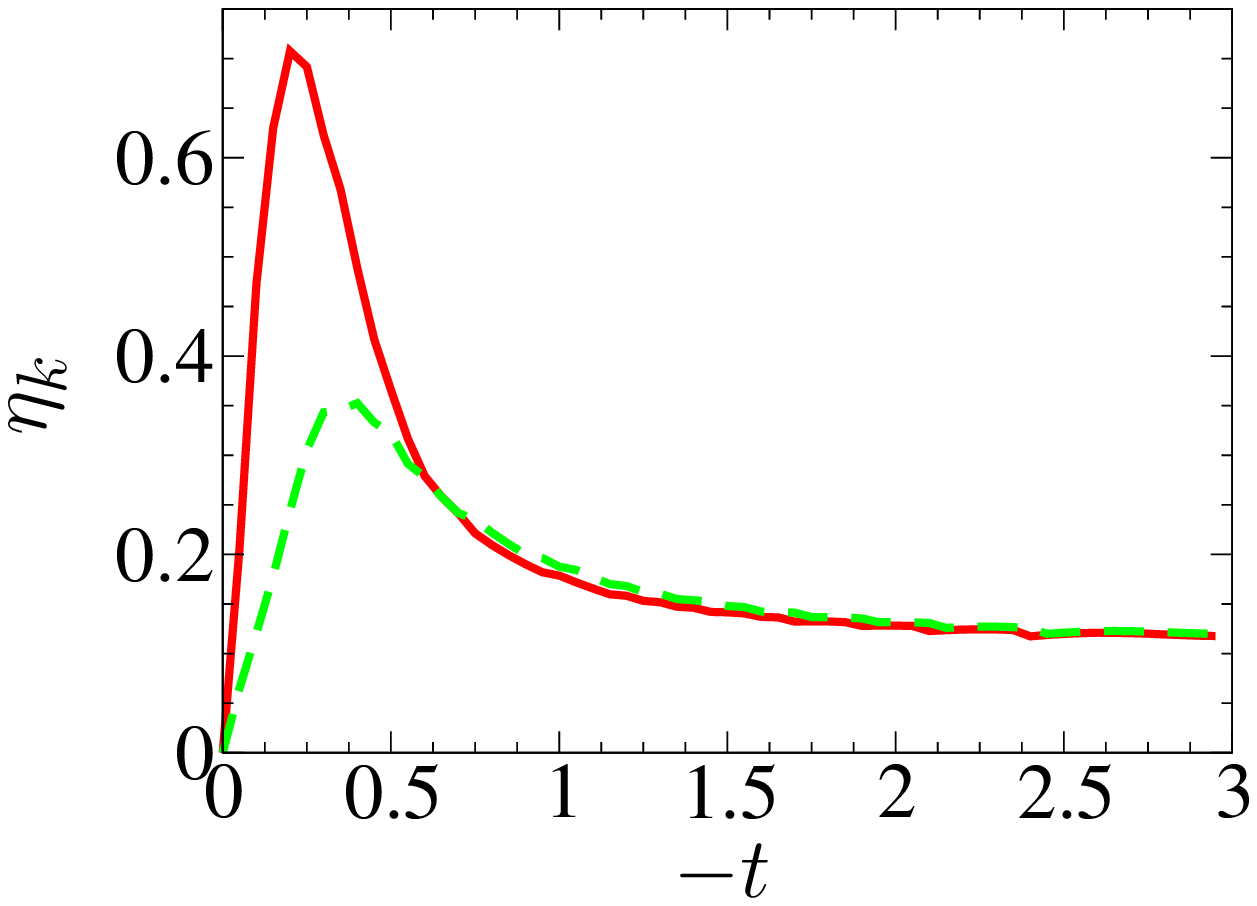}}
\caption{(Color online) Left panel: $Z_k$ vs $-t=\ln(\kin/k)$ obtained with $\rho_{0,k}$ (red solid line) and $\bar\rho_{0,k}$ (green dashed line) ($d=3$). Right panel: Anomalous dimension $\eta_k=-k\dk \ln Z_k$ obtained with $\rho_{0,k}$ (red solid line) and $\bar\rho_{0,k}$ (green dashed line).} 
\label{fig_Z_k} 
\end{figure}

The change of variables (\ref{tilde_def_new}) leads to the flow equations
\begin{align}
k\dk \tilde U_k ={}& \left(-d+\frac{k\dk g_2}{g_2}\right) \tilde U_k \nonumber \\ 
& + \left( d-2+\eta_k-\frac{k\dk g_1}{g_1}\right)\tilde\rho \tilde U_k' \nonumber \\ & + \frac{g_2}{1+\frac{g_1}{g_2}(\tilde U'_k+2\tilde\rho \tilde U''_k)} [(2-\eta_k)I_1 + \eta_k I_2] 
\label{flow_eq_1a}
\end{align} 
and
\beq
\eta_k = 4 \frac{g_1^3}{g_2^2} \tilde\rho \bigl[3\tilde U_k''+2\tilde\rho \tilde U_k''' \bigr]^2 \frac{I_3+\delta_{d,2}/(8\pi)}{\Bigl[1 + \frac{g_1}{g_2}(\tilde U_k'+2\tilde\rho \tilde U''_k)\Bigr]^4} , 
\label{flow_eq_1b}
\eeq
where 
\begin{align}
I_1 &= \frac{k^{-d}}{2} \int_\q \theta\bigl(\eps_k-\eps_0(\q)\bigr) , \nonumber \\
I_2 &= \frac{k^{-d}}{2} \int_\q \frac{\eps_0(\q)}{\eps_k} \theta\bigl(\eps_k-\eps_0(\q)\bigr) , \label{I_def} \\
I_3 &= \frac{k^{2-d}}{4} \int_\q \frac{\eps_0(\q) \partial^2_{q_x} \eps_0(\q) -[\partial_{q_x} \eps_0(\q)]^2}{\eps_0(\q)^2} \theta\bigl(\eps_k-\eps_0(\q)\bigr) \nonumber 
\end{align}
(see Appendix \ref{app_eta}). In Eq.~(\ref{flow_eq_1b}), the rhs should be evaluated at $\tilde\rho_{0,k}$ or $\tilde{\bar{\rho}}_{0,k}$. For $d=3$ and $n>3$, the numerical solution of the flow equations (\ref{flow_eq_1a},\ref{flow_eq_1b}) is stable for $-t_c\geq 2$. The results described below are obtained for $n=4$ and $t_c=-3$. We find $T_c=0.8 \,T_c^{\rm MF}$ when $Z_k$ is defined with respect to the minimum $\rho_{0,k}$ of $U_k$, and $T_c=0.74 \,T_c^{\rm MF}$ if we use the minimum $\bar\rho_{0,k}$ of $\bar U_k$. The result is not as accurate as in the LPA (as anticipated above; see the discussion following Eq.~(\ref{Z_def})). Nevertheless, with $\bar\rho_{0,k}$ (the only case we discuss in the following), it remains within 2 percent of the exact result $T_c^{\rm exact}=0.752 \,T_c^{\rm MF}$. The flow of $Z_k$ is shown in Fig.~\ref{fig_Z_k}. For $-t<1$, $Z_k$ does not differ significantly from $A_k$ (Fig.~\ref{fig_A_k}). In this regime, only the first harmonic (\ie $\cos q_\nu$) is expected to vary and therefore contribute to $Z_k$. For $-t>1$, the renormalization of higher-order harmonics makes $Z_k$ deviate from $A_k$. While $A_k$ saturates to $\sim 1.19$, $Z_k\sim k^{-\eta^*}$ diverges with an exponent given by the anomalous dimension $\eta^*=\lim_{k\to 0}\eta_k$. $\eta^*\simeq 0.1$ is a poor estimate of the exact result $\eta^*\simeq 0.036$ but agrees with previous estimates based on the LPA'~\cite{Dupuis08}. A $\rho$ dependence of $Z_k$ is expected to improve the value of $\eta^*$~\cite{Canet03}.

\subsection{Comparison with HRT} 
\label{subsec_hrt} 

Our approach bears similarities with the Hierarchical Reference Theory (HRT) of fluids~\cite{Parola84,Parola85,Parola86,Ionescu07,Parola09} (for a review, see Ref.~\cite{Parola95}). The HRT is based on an exact treatment of short-distance (hard-core) interactions supplemented by a RG analysis of long-range interactions. The HRT also applies to classical spin models: as in the lattice NPRG, it starts from the local theory (decoupled sites) and takes into account the intersite coupling in a RG approach. Although the final results are very similar to those we have obtained in Secs.~\ref{subsec_spin_lpa} and \ref{subsec_A_k}, the HRT nevertheless differs from the lattice NPRG in some technical aspects, \eg the choice of the cutoff function and the way degrees of freedom are progressively integrated out~\cite{Parola95,Pini93}. Because it was first developed in the context of liquid state theory, the connection between HRT and the more standard formulation of the NPRG~\cite{Berges02} is not always obvious (for a discussion of the relation between HRT and RG, see Refs.~\cite{Caillol06,Caillol09}). By contrast, the lattice NPRG is formulated in the usual language of statistical field theory. The various improvements over the LPA known for continuum models can then be easily implemented in the lattice NPRG. In Sec.~\ref{subsec_lpap}, we have discussed one of these improvements, the LPA', which allows to compute the anomalous dimension $\eta$.

\section{BKT transition in the 2D XY model}
\label{sec_kt}

The NPRG approach to the continuum O(2) (linear) model reproduces most of the universal properties of the BKT transition in two dimensions~\cite{Graeter95,Gersdorff01}. In particular, one finds a value $\tilde\rho_0^*$ of the dimensionless order parameter (the spin-wave ``stiffness'') such that the $\beta$ function $\beta(\tilde\rho_{0,k})=k\dk \tilde\rho_{0,k}$ nearly vanishes for $\tilde\rho_{0,k}>\tilde\rho_0^*$, which reflects the existence of a line of quasi-fixed points. In this low-temperature phase, after a transient regime, the running of the stiffness $\tilde\rho_{0,k}$ becomes very slow, which implies a very large, although not strictly infinite, correlation length $\xi$. The anomalous dimension $\eta_k$ depends on the (slowly varying) stiffness and takes its largest value when the system crosses over to the disordered regime ($\tilde\rho_{0,k}\sim \tilde\rho_0^*$ and $k\sim \xi^{-1}$). When $\tilde\rho_{0,k}<\tilde\rho_0^*$, the essential scaling $\xi\sim e^{a/(\tilde\rho_0^*-\tilde\rho_{0,k})^{1/2}}$ of the correlation length is reproduced~\cite{Gersdorff01}.

In this section, we apply the lattice NPRG to the two-dimensional XY model defined by the Hamiltonian
\beq
H = -\frac{J}{T} \sum_{\mean{\r,\r'}} \S_\r \cdot \S_{\r'} , 
\eeq
where $\S_\r=(\cos\theta_\r,\sin\theta_\r)$ is a 2D classical spin of unit length. Up to a multiplicative constant, the initial value of the partition function $Z_{\kin}[\h]=\prod_\r z(h_\r)$ ($h_\r=|\h_\r|$) is determined by the partition function of a single site, 
\beq
z(h) = \int_0^{2\pi} \frac{d\theta}{2\pi} e^{h\cos\theta} = I_0(h). 
\eeq
The magnetization points along the applied field with an amplitude
\beq
m(h) = \frac{\partial}{\partial h} \ln z(h) = \frac{I_1(h)}{I_0(h)} ,
\label{m_def}
\eeq
where $I_0(h)$ and $I_1(h)$ are modified Bessel functions. Contrary to the Ising model, the function $h(m)$ obtained by inverting (\ref{m_def}) must be computed numerically. For $k=\kin$, the average effective action takes the form
\beq
\Gamma_{\kin}[\m] = \sum_\r U_{\kin}(\rho_\r) + \half \sum_\q \eps_0(\q) \m_{-\q} \cdot\m_{\q} 
\eeq
($\rho_r=\m_\r^2/2$) with
\beq
U_{\kin}(\rho) = - \ln z(h) + h\sqrt{2\rho} - 8\eps_0\rho .
\eeq
Expanding $I_0(h)$ and $I_1(h)$ for small $h$, one finds $m=h/2+\calO(h^3)$ and in turn 
\beq
U_{\kin}(\rho) = 2\rho(1-4\eps_0) + \calO(\rho^2) . 
\eeq
This yields the transition temperature $T_c^{(\kin)}=4J$, which differs from the mean-field result $T_c^{\rm MF}=2J$ for reasons explained in Sec.~\ref{sec_classical_spin}. 

\begin{figure}
\includegraphics[width=5.5cm,clip]{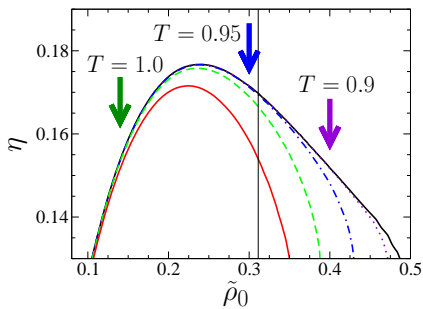}
\caption{(Color online) Flow trajectories $(\tilde\rho_0,\eta)$ for the two-dimensional XY model. $T_c/J=1.05$ (red solid line), 1 (green dashed line), 0.95 (blue dash-dotted line), 0.9 (dotted purple line) and 0.85 (black solid line). The arrows indicate the merging points with the line of quasi-fixed points. The vertical line shows the value of $\tilde\rho_0^*$.}
\label{fig_kt}
\end{figure}

In the LPA', the flow equations read
\begin{align} 
k\dk \tilde U_k ={}& \left(-2+\frac{k\dk g_2}{g_2}\right) \tilde U_k + \left( \eta_k-\frac{k\dk g_1}{g_1}\right) \tilde\rho \tilde U_k' \nonumber \\ & + \left( \frac{g_2}{1+\frac{g_1}{g_2}(\tilde U'_k+2\tilde\rho \tilde U''_k)} + \frac{g_2}{1+\frac{g_1}{g_2}\tilde U'_k} \right) \nonumber \\ & \times [(2-\eta_k)I_1 + \eta_k I_2] ,
\end{align} 
and
\begin{align}
\eta_k ={}& 8 \frac{g_1^3}{g_2^2} \tilde\rho (U_k'')^2 \Bigl[1 + \frac{g_1}{g_2}(\tilde U_k'+2\tilde\rho U''_k)\Bigr]^{-2} \Bigl[1 + \frac{g_1}{g_2}\tilde U_k'\Bigr]^{-2} \nonumber \\ & \times \left(I_3 + \frac{1}{8\pi} \right) ,
\label{eta_def_1} 
\end{align}
where we evaluate the rhs in (\ref{eta_def_1}) at the minimum (which we denote by $\tilde\rho_{0,k}$ for simplicity) of the potential $\tilde{\bar{U}}_k(\tilde\rho)$ [Eq.~(\ref{U_bar_def})] (see Appendix~\ref{app_eta}). $I_1$, $I_2$ and $I_3$ are defined in (\ref{I_def}). We use the change of variables (\ref{tilde_def_new}) with $n=3$ and $t_c=-2$. 

The flow trajectories in the plane $(\tilde\rho_0,\eta)$ are shown in Fig.~\ref{fig_kt}. The flow diagram is reminiscent of the results obtained in the continuum 2D O(2) model~\cite{Graeter95,Gersdorff01}. At low temperature ($T\lesssim J$), the trajectories join a line of quasi-fixed points. The value of $\tilde\rho_0$ at the merging point depends on the temperature. The critical temperature of the BKT transition is defined by the trajectory for which the merging point corresponds to $\tilde\rho_0^*$. A precise determination of the value of $\tilde\rho^*_0$ (which can be obtained by fitting the beta function $k\partial_k\tilde\rho_{0,k}$ to $\frac{1}{\nu}(\tilde\rho_{0,k}-\tilde\rho^*_0)^{3/2}$) is however not possible in the LPA' as it requires the full $\calO(\partial^2)$ expansion of the effective action in the continuum limit $k\ll 1$~\cite{Gersdorff01}. Since for the long-distance properties of the XY model the lattice should not matter, we make the assumption that the ratio $\tilde\rho_0^*/\tilde\rho_{\rm max}$ ($\tilde\rho_{\rm max}$ is the value of $\tilde\rho_0$ for which $\eta$ is maximum) takes the same value in the XY and continuum O(2) models. We can then deduce the value of $\tilde\rho^*_0$ from the results of Ref.~\onlinecite{Gersdorff01}. Although this determination of $T_c$ is clearly approximate, we can nevertheless conclude from our results that $0.9<T_c/J<1$. This estimate should be compared to the exact result $T_c^{\rm exact}=0.89\,J$ obtained from Monte-Carlo calculations~\cite{Olsson95} and the mean-field expression $T_c^{\rm MF}=2J$. Note that the relative error on $T_c$ is of the same order of magnitude as in the 2D Ising model (Sec.~\ref{subsec_spin_lpa}).

\section{Conclusion}

We have proposed a new implementation of the NPRG, which takes as a reference system the local limit of decoupled sites rather than a system where fluctuations are frozen. The lattice NPRG captures both local and critical fluctuations in a non-trivial way. For a lattice field theory and classical spin models, the LPA is sufficient to compute non-universal quantities (transition temperature and magnetization) to an accuracy of the order of 1 percent. We have also discussed an approximation (the LPA') which goes beyond the LPA and allows to compute the anomalous dimension $\eta$. 

A new NPRG scheme has been recently proposed by Blaizot, M\'endez-Galain and Wschebor (BMW)~\cite{Blaizot06,Benitez09}. The BMW approach relies on approximate flow equations for the effective potential $U_k$ and the two-point vertex $\Gamma^{(2)}_k$. Contrary to the LPA and the LPA', it keeps the full momentum dependence of the two-point vertex. We believe that the BMW scheme provides the natural framework to go beyond the LPA in the lattice NPRG.

There are many theoretical methods where the idea to use a reference system which includes short-range fluctuations is central. These methods are usually based on ``cluster'' approaches where one solves exactly (usually numerically) the model on a single cluster (possibly a single site)~\cite{note6} and includes the coupling between clusters by means of a perturbative calculation, a self-consistent condition, etc. The cluster approaches include the correlated cluster mean-field theory in classical spin models (see \eg Ref.~\cite{Yamamoto09} and references therein), the dynamical mean-field theory~\cite{Metzner89,Georges96} (DMFT) and its extensions (cellular DMFT~\cite{Hettler98,Kotliar01}), the cluster perturbation theory~\cite{Gros93,Senechal00}, the variational cluster approach~\cite{Potthoff03c} and the self-energy functional theory~\cite{Potthoff03a}. These approaches describe exactly the local fluctuations but struggle to take into account low-energy (collective) fluctuations which become very important near a phase transition or in low dimensions. 

In this context the lattice NPRG may be seen as a step towards a theory including both local and critical fluctuations in strongly-correlated systems. While we have only discussed classical models, the lattice NPRG can be easily applied to interacting boson systems (Bose-Hubbard model)~\cite{note4}.

\begin{acknowledgments}
We would like to thank B. Delamotte, M.-L. Rosinberg, G. Tarjus and J.-M. Caillol for discussions and/or comments on a preliminary version of the manuscript.
\end{acknowledgments}

\appendix

\section{Anomalous dimension} 
\label{app_eta} 

In this Appendix, we compute the anomalous dimension for the Ising model. From the definition (\ref{Z_def}) and the flow equation (\ref{flow_eq}), one obtains the following expression of the (running) anomalous dimension $\eta_k=-k\dk\ln Z_k$,
\beq
\eta_k = \frac{\gamma_3^2}{2Z_k\eps_0} \int_\q \partial_{q_x} G_k(\q)\partial_{q_x} [k\dk R_k(\q)G_k(\q)^2] ,
\label{app4}
\eeq
where $\gamma_3$ is defined after (\ref{Ak_flow}). To proceed further, we use
\beq
\begin{split}
R &= Z\eps_k y r , \\ 
k\dk R &= -(\eta r+2yr') Z\eps_k y , \\
\partial_{q_x} R &= Z(r+yr') \partial_{q_x} \eps , \\
\partial_{q_x} k\dk R &= - Z[\eta_k r+(\eta+4)yr'+2y^2r''] \partial_{q_x} \eps , \\
\partial_{q_x} G &= -G^2 Z(1+r+yr')\partial_{q_x} \eps , 
\end{split}
\eeq
where $r\equiv r(y)=\theta(1-y)(1-y)/y$, $\eps\equiv \eps_0(\q)$, $G\equiv G(\q)$ and $y=\eps/\eps_k$. To alleviate the notations we now drop the $k$ index. The product of $\partial_{q_x}G\propto (1+r+yr')=\theta(y-1)$ and $\partial_{q_x} k\dk R$ gives zero except for possible singular contributions at $y=1$ coming from $r''r$ and $r''r'$. Thus equation (\ref{app4}) simplifies into 
\beq
\eta = \frac{Z\gamma_3^2}{\eps_0} \int_\q y^2r''(1+r+yr')G^4(\partial_{q_x}\eps)^2 .
\eeq
By an integration by part we obtain
\begin{align}
& \int_\q y^2r''(1+r)G^4(\partial_{q_x}\eps)^2 \nonumber \\ 
={}& \eps_k \int_\q y^2 (1+r)G^4 \partial_{q_x}\eps\partial_{q_x}r' \nonumber \\ 
={}& - \eps_k \int_\mu^\pi \frac{dq_x}{\pi} \int_{\q_\perp} \theta(\q_\perp^2-\mu^2)
r' \partial_{q_x} \bigl[ y^2(1+r)G^4(\partial_{q_x}\eps)^2 \bigr] \nonumber \\ & 
+\frac{\eps_k}{\pi} \int_{\q_\perp} \theta(\q_\perp^2-\mu^2) y^2 (1+r)G^4(\partial_{q_x}\eps)r'\bigr|_{q_x=\mu}^{q_x=\pi} , 
\label{app5}
\end{align} 
where $\q_\perp=(q_y,q_z,\cdots)$. Note that we have regularized the integrals near $\q=0$ ($\mu\to 0^+$). To compute the first integral in (\ref{app5}), $I_1$, we remark that if $\partial_{q_x}$ acts on $G^4$ then the integrand vanishes, so that 
\beq
I_1 = \frac{\bar G^4}{(Z\eps_k)^4} \int_\q \frac{\theta(1-y)}{y^2} \bigl[ \eps(\partial^2_{q_x}\eps)+(\partial_{q_x} \eps)^2 \bigr] , 
\eeq
where $\bar G=Z\eps_k/(Z\eps_k+U'(\rho_0)+2\rho_0U''(\rho_0))$. The second contribution in (\ref{app5}) reads
\beq
I'_1 = \frac{\bar G^4}{\pi Z^4\eps_k^3} \int_{\q_\perp} \theta(\q_\perp^2-\mu^2) \frac{\theta(1-y)}{y} (\partial_{q_x}\eps)\bigl|_{q_x=\mu} , 
\eeq
since $\partial_{q_x}\eps = 2\eps_0\sin(q_x)$ vanishes for $q_x=\pi$. For $\mu\to 0^+$, we obtain
\beq
I'_1 = \frac{2\bar G^4}{\pi Z^4\eps_k^2} \int_{\q_\perp} \theta(\q_\perp^2-\mu^2)\theta(1-y) \frac{\mu}{\q_\perp^2+\mu^2} ,
\eeq
where the Lorentzian $\mu/(\q_\perp^2+\mu^2)$ acts as a delta function $\sim \delta(q_\perp)$. Thus the integral vanishes for $d>2$ and takes the value $\bar G^4/(2\pi Z^4\eps_k^2)$ for $d=2$. By a similar reasoning, we find
\begin{multline}
\int_\q y^3r''r'G^4(\partial_{q_x}\eps)^2  \\ = - \frac{\bar G^4}{2(Z\eps_k)^4} \int_\q \frac{\theta(1-y)}{y^2} \bigl[\eps(\partial^2_{q_x}\eps) + 3(\partial_{q_x}\eps)^2 \bigr] - \half I'_1 . 
\end{multline}
We deduce 
\beq
\eta = \frac{\gamma_3^2\bar G^4}{2\eps_0Z^3\eps_k^2} \left\lbrace \int_\q \frac{\theta(\eps_k-\eps)}{\eps^2} \bigl[\eps(\partial^2_{q_x}\eps) - (\partial_{q_x}\eps)^2 \bigr] + \frac{\delta_{d,2}}{2\pi} \right\rbrace .
\label{app6} 
\eeq
In the continuum limit, one recovers the known results of the LPA'. In particular, when $U(\rho)=\frac{\lambda}{2}(\rho-\rho_0)^2$ is truncated to second order in $\rho$, 
\beq
\eta = 72 \frac{v_d}{d} \frac{\tilde\rho_0\tilde\lambda^2}{(1+2\tilde\rho_0\tilde\lambda)^4} . 
\eeq
Note that the last term in (\ref{app6}) ensures that $\eta$ is a continuous function of $d$. 

For the XY model, a similar calculation leads to (\ref{eta_def_1}).

%\bibliography{/users/lptl/dupuis/publi/BIB/nprg.bib,/users/lptl/dupuis/publi/BIB/bosons.bib,/users/lptl/dupuis/publi/BIB/sft.bib,/users/lptl/dupuis/publi/BIB/stat_phys.bib}

\begin{thebibliography}{43}
\bibitem{note2} When $k\ll 1$, we can use $\eps_0(\q) \simeq \eps_0\q^2$ and 
$\calD(\eps) \simeq 2v_d \eps^{d/2-1}/\eps_0^{d/2}$ where $v_d^{-1}=2^{d+1}\pi^{d/2}\Gamma(d/2)$. Eq.~(\ref{U_flow_1}) then reduces to the usual LPA equation derived in the continuum limit of the model (\ref{action})~\cite{Berges02}.
\bibitem[{not({\natexlab{d}})}]{note7}
\bibinfo{note}{We omit a factor of $T$ in the definition of the
  susceptibility.}
\bibitem[{not({\natexlab{e}})}]{note6}
\bibinfo{note}{Note that one could also take a system of decoupled clusters as
  the reference state in the lattice NPRG.}
\end{thebibliography}

\end{document}